\documentclass[11pt,a4paper]{article}
\usepackage{jheppub}
\usepackage{amsmath}
\usepackage[most]{tcolorbox}
\usepackage{dsfont}
\usepackage{ulem}
\usepackage{natbib}
\usepackage{xcolor}
\usepackage[hang,flushmargin]{footmisc}
\usepackage{tikz-cd}
\usepackage{enumitem}
\usepackage{amsfonts,amssymb, amscd,amsmath,latexsym,amsbsy,bm}
\usepackage{stmaryrd}
\usepackage{todonotes}
\usepackage{float}

\makeatletter\renewcommand{\@biblabel}[1]{#1.}\makeatother

\newtcolorbox{empheqboxed}{colback=gray!20, 
 colframe=white,
 width=\textwidth,
 sharpish corners,
 top=0mm, % default value 2mm
 bottom=0pt
}

\title{Lens Partition Functions and Integrability Properties }

\author{   Mustafa Mullahasanoglu$^{1}$ and Nuri Tas$^1$}
\affiliation{

$^1$ Department of Physics, Bogazici University,\\ 34342 Bebek, Istanbul, Turkey\\[-0.4cm]

}

\emailAdd{mustafa.mullahasanoglu@boun.edu.tr}
\emailAdd{nuri.tas@boun.edu.tr}

\abstract{We study lens partitions functions for the three-dimensional $\mathcal N=2$ supersymmetric gauge theories on $S_b^3/\mathbb{Z}_r$. We consider an equality as a new hyperbolic hypergeometric solution to the star-star relation via the gauge/YBE correspondence. The correspondence allows the construction of integrable lattice spin models of statistical mechanics by the use of integral identities.
Additionally, we obtain new hyperbolic hypergeometric integral identities of gauge theories.
}


\begin{document}
\maketitle
\flushbottom
%%%%%% end of title page %%%%%%

\section{Introduction}

There are many solutions (see \cite{Baxter:1982zz} and references therein) to the two-dimensional Ising model. 
Each solution has an interesting method or a connection to diverse studies.
One of the most interesting solutions of the Ising model is the method of transfer matrices. In this method, there is a condition of the star-triangle relation  \cite{Baxter:1982zz} satisfying the commutation relation of transfer matrices containing spin interaction information at every element of the matrix. That is, the star-triangle relation gives the commutativity property of two transfer matrices to ensure integrability property of the model. 
Namely, the star-triangle relation can be considered as the simplest version of the Yang-Baxter equation (YBE).

There is also a star-star relation which is an integrability condition for spin models and can be obtained in the existence\footnote{There are also examples for that a model has a star-star relation but has not a star-triangle relation, see \cite{baxter:1997ssr}.} of the star-triangle relation. However, the star-star relation is the simpler condition for lattice spin models since one can have an integrable spin model in which there is not a star-triangle relation but the Boltzmann weight of the model satisfy a star-star relation \cite{baxter:1997ssr, Bazhanov:2013bh}.
We also study interaction-round-a-face (IRF) spin models where four spins interact over a face and another version of the YBE for the integrability condition of the IRF model is called IRF-type YBE. Another property of the star-star relation is that a star-star relation \cite{baxter:1997ssr} is enough to write an IRF-type YBE.

We note that the star-triangle relation and the star-star relation from the Ising-like model enable us to construct factorized IRF-type models. Here, factorized (\ref{factorized}) means that an IRF of four spins can be written by four-edge interactions.  
 \begin{figure}[tbh]
\centering
\begin{tikzpicture}[scale=1]
\filldraw[fill=black,draw=black] (-5.55,-0.35) 

node[] {\color{black}  {\tiny $SUSY| ~G:SU(2)~F:SU(6)$} };
\filldraw[fill=black,draw=black] (-5.55,-0.8) 

node[] {\color{black}  {\tiny YBE$|$ ~Star-triangle relation\cite{Gahramanov:2016ilb}} };

\draw[-,thick] (-8,0)--(-3,0);
\draw[-,thick] (-8,-1)--(-3,-1);
\draw[-,thick] (-8,0)--(-8,-1);
\draw[-,thick] (-3,-1)--(-3,0);

\filldraw[fill=black,draw=black] (4.55,-0.25) 

node[] {\color{black}  {\tiny $SUSY|~ G:SU(2)$} };
\filldraw[fill=black,draw=black] (4.55,-0.5) 

node[] {\color{black}  {\tiny $ F:SU(8)$ } };
\filldraw[fill=black,draw=black] (4.55,-0.8) 

node[] {\color{black}  {\tiny YBE$|$~ Star-star relation~\ref{newssr} } };

\draw[-,thick] (2,0)--(7,0);
\draw[-,thick] (2,-1)--(7,-1);
\draw[-,thick] (2,0)--(2,-1);
\draw[-,thick] (7,-1)--(7,0);

\filldraw[fill=black,draw=black] (-5.55,-3.35) 

node[] {\color{black}  {\tiny $SUSY|~ G:U(1)~ F:SU(3)\times SU(3)$} };
\filldraw[fill=black,draw=black] (-5.55,-3.8) 

node[] {\color{black}  {\tiny YBE$|$~ Star-triangle relation\cite{Bozkurt:2020gyy}} };

\draw[-,thick] (-8,-3)--(-3,-3);
\draw[-,thick] (-8,-4)--(-3,-4);
\draw[-,thick] (-8,-3)--(-8,-4);
\draw[-,thick] (-3,-4)--(-3,-3);

\filldraw[fill=black,draw=black] (4.55,-3.25) 

node[] {\color{black}  {\tiny $SUSY|~ G:U(1)$} };
\filldraw[fill=black,draw=black] (4.55,-3.5) 

node[] {\color{black}  {\tiny $ F:SU(4)\times SU(4)$ } };
\filldraw[fill=black,draw=black] (4.55,-3.8) 

node[] {\color{black}  {\tiny YBE$|$~ Star-star relation\cite{Catak:2021coz}} };

\draw[-,thick] (2,-3)--(7,-3);
\draw[-,thick] (2,-4)--(7,-4);
\draw[-,thick] (2,-3)--(2,-4);
\draw[-,thick] (7,-4)--(7,-3);

\draw[->,dashed,violet] (-2.5,-0.55).. controls (-0.5,0.3) .. (1.5,-0.55);

\filldraw[fill=black,draw=black] (-0.5,0.3) 

node[] {\color{black} derivation};

\draw[->,dashed,violet] (-2.5,-3.55).. controls (-0.5,-4.3) .. (1.5,-3.55);

\filldraw[fill=black,draw=black] (-0.5,-4.3) 

node[] {\color{black} derivation};

\draw[->,thick,violet] (-5.55,-1.2).. controls (-5.55,-1.3) .. (-5.55,-2.8);

\filldraw[fill=black,draw=black] (-5.55,-2) 

node[] {\color{black} $gauge~ symmetry~ breaking$};

\draw[->,thick,violet] (4.55,-1.2).. controls (4.55,-1.3) .. (4.55,-2.8);

\filldraw[fill=black,draw=black] (4.55,-2) 

node[] {\color{black} $gauge~ symmetry~ breaking$};
\label{four}
\end{tikzpicture}
\caption{In the present study, we complete the scheme.}
\end{figure}
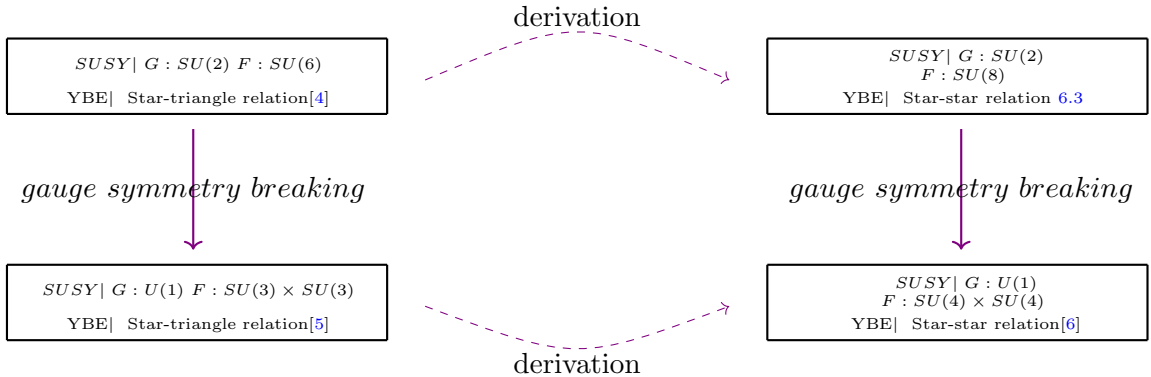

In the last decade, there has been a lot of works constructing solutions to the star-triangle and the star-star relations with a remarkable correspondence between the duality of supersymmetric gauge theories and the Yang-Baxter equation, which is called the "gauge/YBE correspondence". For more examples about the correspondence and comprehensive explanations, see, e.g. \cite{Gahramanov:2017ysd, Yamazaki:2018xbx}. The correspondence enables us to write the star-triangle relation \cite{Spiridonov:2010em,Kels:2015bda,Gahramanov:2015cva,Gahramanov:2016ilb,Bozkurt:2020gyy} or the star-star relations \cite{Yamazaki:2013nra,Kels:2017vbc,Catak:2021coz} of exactly solvable lattice spin models in statistical mechanics with the interpretation of integral identities resulting from the equality of supersymmetric partition functions. Simply, the duality of two supersymmetric gauge theories gives us an integral identity and this identity can be interpreted as a star-triangle relation or a star-star relation. 

On the other hand, non-perturbative supersymmetric gauge theories have various aspects in mathematical structure, see, e.g. \cite{Gahramanov:2015tta, Gahramanov:2022qge, Spiridonov:2019kto, Tachikawa:2017byo}.
Like our results, especially in the last section, it can be seen as the source of new identities for hypergeometric functions \cite{Spiridonov:2009za,Spiridonov2014,Krattenthaler:2011da, Gahramanov:gka,Dolan:2011rp}.

Here, as the novel part of the present work, we obtain a new hyperbolic hypergeometric solution (\ref{newssr}) to the star-star equation by using the gauge/YBE correspondence. From a mathematical perspective, the importance of a new solution to the star-star equation is presented in \cite{Bazhanov:2011mz, Bazhanov:2013bh, Kels:2017toi, Rains:2003} as a sum/integral identity and related transformation and $r=1$ case of the new solution is discussed in \cite{Sarkissian:2020ipg} as a transformation. 
Another novelty of our study is the presentation of setting a bridge between two solutions to the star-star equation via the gauge symmetry-breaking method.

We also obtain the solution to the star-star equation (\ref{u1ssr}) which was obtained in \cite{Catak:2021coz} by applying the gauge symmetry breaking method, firstly presented in \cite{Spiridonov:2010em}, to the novel solution. In the supersymmetric field theory side, the gauge group of $\mathcal{N}=2$ supersymmetric dual pair on $\mathcal{S}^3_b/\mathbb{Z}_r$ will be reduced from $SU(2)$ to the $U(1)$ gauge group. Thus, we complete the scheme shown pictorially in Fig.\ref{four}, between the supersymmetric theories and the spin models in statistical mechanics, namely, the method connects star-star equations as presented in \cite{Bozkurt:2020gyy} for the star-triangle equations. Thus we also construct the star-star relation for the generalized Faddeev-Volkov model \cite{Catak:2021coz} by breaking the $SU(2)$ gauge symmetry group to $U(1)$.

As a final part of the study, we present new hyperbolic hypergeometric integral identities in which the case $r=1$ (partition functions on $S_b^3$) agree with the results of \cite{Sarkissian:2020ipg}\footnote{Note that our resulting dual theories need superpotential \cite{Khmelnitsky:2009vc} constructions since the theories are dual under the presence of some particular superpotentials.}.

The rest of the article has the following structure. In Section 2, we briefly touch on mathematical tools and introduce them. Section 3 contains the $\mathcal{N}=2$ on $S^3_b/\mathbb{Z}_r$ partition functions of supersymmetric gauge theory. The integrability of the Ising-like models and IRF-type models is discussed in Section 4. Quiver notations of the dualities are considered in Section 5. Solutions to the star-star equation and the transition from one solution to another are discussed in Section 6. Some extra results are performed in Section 7.

\section{Notations}
	In this work, we mostly use the hyperbolic gamma function, which is a variant of Faddeev's non-compact quantum dilogarithm \cite{van2007hyperbolic, Andersen:2014aoa}.

	\par The first definition of the hyperbolic gamma function can be given as the infinite product representation
%tanım hatalı, spiridonova göre yeniden yaz.
	\begin{align}
	\gamma^{(2)}(z;\omega_{1},\omega_{2})=e^{\frac{\pi i}{2}B_{2,2}(z;\omega_{1},\omega_{2})}\frac{(e^{2\pi i\frac{z}{\omega_{2}}}\tilde{q};\tilde{q})_\infty}{(e^{2\pi i\frac{z}{\omega_{1}}};q)_\infty} \; ,
	\end{align}
	where the parameters are defined as $\tilde{q}=e^{2\pi i \omega_{1}/\omega_{2}}$ and $q=e^{-2\pi i \omega_{2}/\omega_{1}}$   and the Bernoulli polynomial is
\begin{equation}
 B_{2,2}(z;\omega_1,\omega_2)=\frac{z^2}{\omega_1\omega_2}-\frac{z}{\omega_1}-\frac{z}{\omega_2}-\frac{\omega_1}{6\omega_2}-\frac{\omega_2}{6\omega_1}+\frac{1}{2}\:,
\end{equation}
with the complex variables $\omega_{1}$, $\omega_{2}$.

The second diagonal Bernoulli polynomial $B_{2,2}(z;\omega_1,\omega_2)$ will be frequently utilized to prove the breakings of the integral identities in this study.

One of the several integral representations for the hyperbolic gamma function, see, e.g. \cite{Faddeev:1995nb,woronowicz2000quantum}, is 
	\begin{align}
	\gamma^{(2)}(z;\omega_{1},\omega_{2})=\exp{\left(-\int_{0}^{\infty}\frac{dx}{x}\left[\frac{\sinh{x(2z-\omega_{1}-\omega_{2})}}{2\sinh{(x\omega_{1})}\sinh{(x\omega_{2})}}-\frac{2z-\omega_{1}-\omega_{2}}{2x\omega_{1}\omega_{2}}\right]\right)} \; ,
	\end{align}
	where $Re(\omega_{1}),Re(\omega_{2})>0$ and $Re(\omega_{1}+\omega_{2})>Re(z)>0$. \\

It should be noted that the asymptotic behaviors of the function for Im$(\omega_1/\omega_2)>0$ happen to be

	\begin{align}
	\begin{aligned}
	\lim_{z\to\infty}e^{\frac{\pi i}{2}B_{2,2}(z;\omega_{1},\omega_{2})}\gamma^{(2)}(z;\omega_{1},\omega_{2})=1\: \: \text{for} \:  \: \arg{\omega_{2}+\pi}>\arg{z}>\arg{\omega_{1}} , \\
	\lim_{z\to\infty}e^{-\frac{\pi i}{2}B_{2,2}(z;\omega_{1},\omega_{2})}\gamma^{(2)}(z;\omega_{1},\omega_{2})=1 \: \: \text{for} \: \: \arg{\omega_{2}}>\arg{z}>\arg{\omega_{1}-\pi} \; .
\label{asymp}
	\end{aligned}
	\end{align}
The reflection property of the hyperbolic gamma function is used throughout the paper
	\begin{align}
\gamma^{(2)}(\omega_{1}+\omega_{2}-z;\omega_{1},\omega_{2})\gamma^{(2)}(z;\omega_{1},\omega_{2})=1 \:,
	\end{align}
	
	and we adopt the following conventional writing
	\begin{align}
\gamma^{(2)}(\pm z;\omega_{1},\omega_{2})=\gamma^{(2)}(z;\omega_{1},\omega_{2})\gamma^{(2)}(-z;\omega_{1},\omega_{2})  \:.
	\end{align}
	
We also introduce the improved double sine function defined in \cite{Nieri:2015yia} 
\begin{align}
 \hat{s}_{b, -m}(z) = e^{i\pi\phi([\![m]\!])}\gamma^{(2)}\left(iz+m\omega_1+\frac{\omega}{2};\omega_1r,\omega\right) 
 \gamma^{(2)}\left(iz+\omega_2(r-m)+\frac{\omega}{2};\omega_2r,\omega\right),
\end{align}
where $\phi([\![m]\!])=\frac{1}{2r}([\![m]\!](r-[\![m]\!])-(r-1)[\![m]\!]^2)$ and $[\![m]\!]_r \in \{0, 1, . . . , r - 1\}$ denotes m modulus r.  

The shorthand notation $\omega=\omega_{1}+\omega_{2}$ will be used for the rest of the paper.

\section{Supersymmetric partition function on \texorpdfstring{$S_b^3/\mathbb{Z}_r$}{Sb3/Zr}}

\subsection{\texorpdfstring{$\mathcal{N}=2$}{N=2} supersymmetry in three-dimensions}

There are four supercharges in three-dimensional $\mathcal{N}=2$  supersymmetric gauge theories. Such theories can be obtained by reducing four-dimensional $\mathcal{N}=1$ theories. The three-dimensional $\mathcal{N}=2$ supersymmetry algebra has also the rotational symmetry of supercharges with $SO(2)$ group as occurring in four-dimensions and $R$-symmetry rotates supercharges.

In the three-dimensional $\mathcal{N}=2$ supersymmetry, the partition function of a gauge theory contains contributions coming from vector and matter multiplets and classical terms. In our discussion, there will be no contribution from classical terms stemming from Chern-Simons and Fayet Iliopoulos terms in classical action. 

Vector potential takes place in the vector multiplet belonging to an adjoint representation of a gauge group $G$ and, in the three-dimensional $\mathcal{N}=2$, there is also real scalar potential living in the adjoint representation of the same gauge group. The vector multiplet contains, in addition to the gauge field and real scalar field, a complex Dirac fermion and auxiliary scalar field. The vector term in the partition function comes from this vector multiplet. 

The matter contribution in the partition function comes from chiral multiplets. In this paper, chiral multiplets belong to the fundamental (or anti-fundamental) representation of the gauge group $G$. 
\subsection{Supersymmetric partition functions on \texorpdfstring{$S_b^3/\mathbb{Z}_r$}{Sb3/Zr}}
Under the given identifications $(x,y) \to (e^{\frac{2\pi i}{r}} x,  e^{-\frac{2\pi i}{r}} y)$, a squashed three-sphere

\begin{equation} 
S_b^3=\{(x,y)\in \mathbb{C}^2| \;\; b^2|x|^2+b^{-2}|y|^2=1\} \;,
\end{equation}

turns to be a squashed lens space $S_b^3/\mathbb{Z}_r$.

A general formula for the partition function on $S_b^3/\mathbb{Z}_r$ is given by
\begin{align} \label{generalZ}
Z & =  \sum_{m}\int d\mu e^{-S_{\rm cl}} Z_{\rm  1-loop} \;,
\end{align}
where $d\mu$ is the integral measure, $Z_{\rm  1-loop}$ consists of vector and matter terms, the summation is over holonomies and finally, $e^{-S_{\rm cl}}$ is a contribution from the classical action of Chern-Simons and Fayet-Iliopoulos terms. 

The measure of integration is defined as
\begin{align} 
d\mu=\frac{ 1}{ |\mathcal{W}|}\prod_{j=1}^{\text{rank G}}\frac{  dz_j}{2\pi ir} \;.
\end{align}

Then one can rewrite the partition function as
\begin{align} 
Z & =  \sum_{m}\int\frac{ 1}{ |\mathcal{W}|}\prod_{j=1}^{\text{rank G}}\frac{  dz_j}{2\pi ir}   \; Z_{\rm vector}[z, m] \; Z_{\rm matter}[z, m] \;,
\end{align}
where the summation consists of holonomies m defined with an integrating gauge field $A_\mu$ over a non-trivial cycle $C$ on the squashed lens space $\mathcal{S}^3_b/\mathbb{Z}_r$ as
\begin{align}
  m \ = \ \frac{r}{2 \pi} \int_C A_\mu dx^\mu \:.
\end{align}
In our case, there will be one variable since the gauge group is $SU(2)$ and the variables have the relation $z_1+z_2=0$.

For theories with non-abelian gauge groups, together with Vandermonde determinant,  $Z_{\text{vector}}$ term \cite{Gahramanov:2016ilb} coming from the one-loop contribution of the vector multiplet is as follows 
\begin{align} \nonumber
Z_{\rm vector} & = \prod_{\alpha}\frac{1}{\hat s_{b, \alpha(m)} 
\left( i\frac{Q}{2}+ \alpha(z)\right)} ,
\end{align}
where the product is taken with the $\alpha$ indices over the positive roots of the gauge group $G$. 

Another term $Z_{\text{matter}}$ \cite{Gahramanov:2016ilb} coming from the one-loop contribution of the chiral multiplets is as the following
\begin{equation}
Z_{\rm matter}=
\prod_{i}\prod_{\rho_i}\prod_{\phi_i}
\hat s_{b,-\rho_i( m)-\phi_i( n)} 
\left( i\frac{Q}{2} (1-\Delta_i)-\rho_i(z)-\phi_i( \Phi)\right),
\end{equation}
where the product $i$ over different chiral multiplets,  $\rho_i$ and $\phi_i$ are for the gauge  and  flavor groups, respectively and they stand as the weights of the representation of the associated groups and $\Delta_i$ labels the Weyl weight of $i$'th chiral multiplet. The squashing parameter (real or phase \cite{Imamura:2011wg, Closset:2012ru}) $b^2=\omega_2/\omega_1$  defines $Q=b+\frac{1}{b}$.

\subsection{\texorpdfstring{$\mathcal{N}=2$}{N=2} Seiberg duality in three-dimensions}
\label{3d}
Seiberg proposed \cite{Seiberg:1994pq} the equivalence of "physics" at the same infrared fixed point of two theories behaving differently at the ultraviolet level. One of the strong pieces of evidence of the existence of dualities is the equality of partition functions of dual theories in infrared fixed point
\begin{equation}
    {\Large \mathcal{Z}_{\textbf{theory A}}} = {\Large \mathcal{Z}_{\textbf{theory B}}} \;.
\end{equation}
In three-dimensional $\mathcal N=2$ theories, the duality argument is shown in \cite{Intriligator:1996ex, Aharony:1997bx}. Three-dimensional dualities have been verified at the level of sphere partition functions (e.g. \cite{Kapustin:2010xq}), squashed sphere partition functions (e.g. \cite{Dolan:2011rp,Gahramanov:gka,Amariti:2015vwa}), superconformal indices (e.g. \cite{Krattenthaler:2011da,Kapustin:2011jm,Gahramanov:2013rda,Gahramanov:2016wxi}), lens partition functions (e.g. \cite{Benini:2011nc, Imamura:2012rq,Imamura:2013qxa}) and so on.

Our discussion on dualities obtained via reductions are derived from a special case of $SP(2N)$ three-dimensional $\mathcal N=2$ supersymmetric duality considered in \cite{Aharony:2013dha}.
\begin{figure}[tbh]
\centering
\begin{tikzpicture}[scale=1]

\filldraw[fill=black,draw=black] (-6.9,-1) 
%circle (1.2pt)
node[] {\color{black}  {\Large  $\mathcal{Z}^{\textbf{A}}_{\mathcal{S}^3_b/\mathbb{Z}_r}$}};

\filldraw[fill=black,draw=black] (-4,-0.75) 
%circle (1.2pt)
node[] {\color{black} $ Gauge=SU(2)$};

\filldraw[fill=black,draw=black] (-4,-1.25) 
%circle (1.2pt)
node[] {\color{black} $ Flavor~ Group:SU(6)$};

\draw[-,dashed,blue] (-6,0)--(-2,0);
\draw[-,dashed,blue] (-6,-2)--(-2,-2);
\draw[-,dashed,blue] (-6,0)--(-6,-2);
\draw[-,dashed,blue] (-2,-2)--(-2,0);

\filldraw[fill=black,draw=black] (-1.6,-1) 
%circle (1.2pt)
node[] {\color{black} \textbf{=}};

\filldraw[fill=black,draw=black] (-0.87,-1) 
%circle (1.2pt)
node[] {\color{black}  {\Large  $\mathcal{Z}^{\textbf{B}}_{\mathcal{S}^3_b/\mathbb{Z}_r}$}};

\filldraw[fill=black,draw=black] (2,-0.75) 
%circle (1.2pt)
node[] {\color{black} $No~Gauge~Symmetry$ };

\filldraw[fill=black,draw=black] (2,-1.25) 
%circle (1.2pt)
node[] {\color{black} $15~ Chiral~ Multiplets$ };

\draw[-,dashed,green] (0,0)--(4,0);
\draw[-,dashed,green] (0,-2)--(4,-2);
\draw[-,dashed,green] (0,0)--(0,-2);
\draw[-,dashed,green] (4,-2)--(4,0);

\end{tikzpicture}
\caption{Seiberg duality and ingredients of dual two theories on $\mathcal{S}^3_b/\mathbb{Z}_r$. }
\label{Seibergduality}
\end{figure}

The lens partition function of \textbf{Theory A} is

\begin{align}
{\Large  \mathcal{Z}^{\textbf{A}}_{\mathcal{S}^3_b/\mathbb{Z}_r}} & = \sum_{m_0=0}^{r-1}\int_{\mathbb{R}}\frac{ dx_0}{2r\sqrt{\omega_1\omega_2}}\;\frac{\prod_{k=1}^{6}\frac{\hat s_{b,-m_0-m_{k}}(x_0+x_{k}+i Q/2)}{\hat s_{b,-m_0+\bar m_{k}}(x_0-x_k-i Q/2)}}{\hat s_{b,-2m_0} 
\left( i\frac{Q}{2}+ 2x_0\right)} ~,
\label{thryA}
\end{align}
where the balancing conditions are $i\sum_{i=1}^6 x_i=Q$, $\sum_{i=1}^6 m_i=0$.
The numerator of the fraction in the integrand of the partition function involves the contribution of chiral multiplets that transform under the fundamental representation of the gauge group and the flavor group, while the denominator contains the contributions of a vector multiplet that transforms as the adjoint representation of the gauge group.

The \textbf{Theory B} will have the following lens partition function

\begin{equation}
{\Large  \mathcal{Z}^{\textbf{B}}_{\mathcal{S}^3_b/\mathbb{Z}_r}}   \ = \ \prod_{1\leq j<k\leq 6} \hat s_{b,-m_{j}-m_{k}}(x_j+x_k+i Q/2) \;,
\label{thryB}
\end{equation}
where the balancing conditions are $i\sum_{i=1}^6 x_i=Q$, $\sum_{i=1}^6 m_i=0$. The lens partition function is simpler and 
the chiral multiplets are in the totally antisymmetric fifteen dimensional tensor representation of the flavor group. 

Instead of writing (\ref{thryA}) and (\ref{thryB}) in terms of generalized double sine function notation used  mostly in the study of supersymmetric partition functions, we write them in terms of the hyperbolic hypergeometric function to be consistent with the studies on integrable spin-lattice models. The exponential factors in double sine function notation cancel each other when we have hyperbolic hypergeometric gamma functions in the identity. Then, supersymmetric duality reads the equality of lens partition functions as

	\begin{align} \nonumber
	\frac{1}{2r\sqrt{-\omega_1\omega_2}} \sum_{y=0}^{[ r/2 ]}\epsilon (y) \int _{-\infty}^{\infty} dz\frac{\prod_{i=1}^6
	\gamma^{(2)}(-i(a_i\pm z)-i\omega_1(u_i\pm y);-i\omega_1r,-i\omega)}
	{\gamma^{(2)}(\pm 2iz\pm i\omega_12y;-i\omega_1r,-i\omega)} \nonumber \\
	\times \frac{\gamma^{(2)}(-i(a_i\pm z)-i\omega_2(r-(u_i\pm y));-i\omega_2r,-i\omega)}
	{\gamma^{(2)}(\pm2iz-i\omega_2(r\pm2y);-i\omega_2r,-i\omega)} \nonumber \\ =\prod_{1\leq i<j\leq 6}\gamma^{(2)}(-i(a_i + a_j)-i\omega_1(u_i + u_j);-i\omega_1r,-i\omega) \nonumber \\ \times \gamma^{(2)}(-i(a_i + a_j)-i\omega_2(r-(u_i + u_j));-i\omega_2r,-i\omega)\; ,
	\label{Sb3Zr}
	\end{align}
where balancing conditions are $\sum_{i=1}^6a_i=\omega_1+\omega_2$ and $\sum_{i=1}^6u_i=0$. The function of $\epsilon(y)$ takes values for all situations $\epsilon(y)=2$ except $\epsilon(0)=\epsilon(\lfloor\frac{r}{2}\rfloor)=1$.

\subsection{Gauge Symmetry Breaking}\label{gsb}

In this section, we explain the work \cite{Bozkurt:2018xno} and give some details of the new result by applying gauge symmetry breaking.	
The idea is first done in \cite{Spiridonov:2010em} (see also \cite{Sarkissian:2018ppc}) for the duality of partition functions on $S_b^3$ and used in \cite{Bozkurt:2020gyy} for the dual theories on $S_b^3/\mathbb{Z}_r$. Both studies perform the way of doing gauge symmetry breaking step by step. 

From a physical point of view, gauge symmetry breaking is adding VEV to two flavor quarks to reduce the gauge group from $SU(2)$ to $U(1)$ and to break the flavor group $SU(3) \times SU(3)$. Then, the duality represented in Fig.\ref{Seibergduality} turns to the following duality represented in Fig.\ref{Seibergduality2}.
\begin{figure}[tbh]
\centering
\begin{tikzpicture}[scale=1]

\filldraw[fill=black,draw=black] (-6.9,-1) 
%circle (1.2pt)
node[] {\color{black}  {\Large  $\mathcal{Z}^{\textbf{A}}_{\mathcal{S}^3_b/\mathbb{Z}_r}$}};

\filldraw[fill=black,draw=black] (-4,-0.75) 
%circle (1.2pt)
node[] {\color{black} $ Gauge=U(1)$};

\filldraw[fill=black,draw=black] (-4,-1.25) 
%circle (1.2pt)
node[] {\color{black} $ Flavor:SU(3) \times SU(3)$};

\draw[-,dashed,blue] (-6,0)--(-2,0);
\draw[-,dashed,blue] (-6,-2)--(-2,-2);
\draw[-,dashed,blue] (-6,0)--(-6,-2);
\draw[-,dashed,blue] (-2,-2)--(-2,0);

\filldraw[fill=black,draw=black] (-1.6,-1) 
%circle (1.2pt)
node[] {\color{black} \textbf{=}};

\filldraw[fill=black,draw=black] (-0.87,-1) 
%circle (1.2pt)
node[] {\color{black}  {\Large  $\mathcal{Z}^{\textbf{B}}_{\mathcal{S}^3_b/\mathbb{Z}_r}$}};

\filldraw[fill=black,draw=black] (2,-0.75) 
%circle (1.2pt)
node[] {\color{black} $No~Gauge~Symmetry$ };

\filldraw[fill=black,draw=black] (2,-1.25) 
%circle (1.2pt)
node[] {\color{black} $9~ Chiral~ Multiplets$ };

\draw[-,dashed,green] (0,0)--(4,0);
\draw[-,dashed,green] (0,-2)--(4,-2);
\draw[-,dashed,green] (0,0)--(0,-2);
\draw[-,dashed,green] (4,-2)--(4,0);

\end{tikzpicture}
\caption{Duality with broken gauge symmetry and ingredients of them on $\mathcal{S}^3_b/\mathbb{Z}_r$. }
\label{Seibergduality2}
\end{figure}

In \textbf{Theory A}, chiral multiplets are belonging to the $SU(3)\times SU(3)$ in two classes: one group transforms in the fundamental representation of the gauge group and another transforms in the anti-fundamental representation. Similar to the previous duality, \textbf{Theory B} is gauge invariant for all physical degrees of freedom and nine chiral multiplets transform in the fundamental representation of the flavor group $SU(3) \times SU(3)$.

The details of gauge symmetry breaking are as follows

\begin{itemize}
 \item Before we get started, we change the boundary of integral (\ref{Sb3Zr}) from $\int _{-\infty}^{\infty}$ to $2\int _{0}^{\infty}$ by the symmetry property of $z$\:.
 \item We replace flavor fugacities $a_i$ to $a_i+\mu$ for $i=\{1,2,3\}$ and $a_i-\mu$ for $i=\{4,5,6\}$,
 \item We change the variable $z$ to $z+\mu$\:.
 \item Finally, we take the limit $\mu\to\infty$ and use the asymptotic behaviour of hyperbolic hypergeometric gamma functions (\ref{asymp}) and we rename the flavor group coefficients as $a_{i+3}\to b_i$ and $u_{i+3}\to v_i$ for $i=\{1,2,3\}$.
\end{itemize}

After doing all calculations, one obtains the following integral identity representing the equality of \textbf{Theory A}(LHS) and \textbf{Theory B}(RHS)
	
\begin{align}
\frac{1}{r\sqrt{-\omega_1\omega_2}}\sum_{y=0}^{[ r/2 ]}\epsilon (y) e^{\frac{\pi iC}{2}}\int _{-\infty}^{\infty} dz\prod_{i=1}^3 & \gamma^{(2)}(-i(a_i-z)-i\omega_1(u_i- y);-i\omega_1r,-i\omega) \nonumber \\
	\times & \gamma^{(2)}(-i(a_i-z)-i\omega_2(r-(u_i- y));-i\omega_2r,-i\omega) \nonumber \\
	\times & \gamma^{(2)}(-i(b_i+z)-i\omega_1(v_i+ y);-i\omega_1r,-i\omega) \nonumber
	\\
	\times & \gamma^{(2)}(-i(b_i+z)-i\omega_2(r-(v_i+y));-i\omega_2r,-i\omega)\nonumber\\
	= \prod_{i,j=1}^3& \gamma^{(2)}(-i(a_i+ b_j)-i\omega_1(u_i+ v_j);-i\omega_1r,-i\omega)\nonumber\\ \times& \gamma^{(2)}(-i(a_i+ b_j)-i\omega_2(r-(u_i+ v_j));-i\omega_2r,-i\omega))\;, \label{u1str}
	\end{align}

with the balancing conditions $\sum_i a_i+b_i=\omega_1+\omega_2$ and $\sum_i u_i+v_i=0$ and where $C=-2y+(u_1+u_2+u_3-v_1-v_2-v_3)$. 

In the case of $r=1$, the duality of partition functions will be on squashed sphere $S_b^3$ \cite{Teschner:2012em,Kashaev:2012cz} and it is discussed for the superconformal indices in \cite{Gahramanov:2013rda,Gahramanov:2014ona,Gahramanov:2016wxi}. 

This result (\ref{u1str}) is also represented as the pentagon identity \cite{Bozkurt:2020gyy} which stands for the basic 2-3 Pachner move \cite{pachner1991pl} for a certain triangulated 3-manifold.
Three-dimensional supersymmetric dualities are also the source for integral pentagon relations, see, e.g. \cite{Dimofte:2011ju,Kashaev:2012cz,Gahramanov:2013rda, Gahramanov:2014ona, Gahramanov:2016wxi, Benvenuti:2016wet, Bozkurt:2018xno, Jafarzade:2018yei}.

\section{The two-dimensional integrable lattice models}

In this section, we consider Ising-like models and the integrability property of these spin-lattice models. We study Ising-like models where there will be two kinds of spin variables $x$ (in $\mathbb{R}$ or its subsets)  and $m$ (in $\mathbb{Z}$ or its subsets) that are continuous and discrete, respectively. We will use the notation $\sigma_i=(x_i,m_i)$ (in $\mathbb{R}\times\mathbb{Z}$ or its subsets) to describe continuous and discrete spin variables.

\subsection{Square lattice model}

The two-dimensional Ising-like model is a nearest-neighbor spin interaction on a square lattice with periodic boundary conditions. 

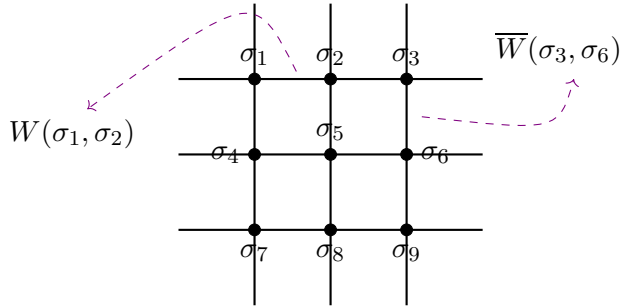
\begin{figure}[tbh]
\centering
\begin{tikzpicture}[scale=1]
\draw[-,thick] (-4,0)--(0,0);
\draw[-,thick] (-4,-2)--(0,-2);
\draw[-,thick] (-4,-1)--(0,-1);
\draw[-,thick] (-3,1)--(-3,-3);
\draw[-,thick] (-2,1)--(-2,-3);
\draw[-,thick] (-1,1)--(-1,-3);

\draw[->,dashed,violet]  (-0.8,-0.5)..  controls (1,-0.7) .. (1.2,0);

\filldraw[fill=black,draw=black] (1,0) 
%circle (1.2pt)
node[above=1.5pt] {\color{black} $\overline{W}(\sigma_3,\sigma_6)$};

\draw[->,dashed,violet]  (-2.45,0.1)..  controls (-3,1.2) .. (-5.2,-0.4);

\filldraw[fill=black,draw=black] (-5.4,-1.1) 
%circle (1.2pt)
node[above=1.5pt] {\color{black} $W(\sigma_1,\sigma_2)$};

\filldraw[fill=black,draw=black] (-3,0) circle (2.2pt)
node[above=1.5pt] {\color{black} $\sigma_1$};
\filldraw[fill=black,draw=black] (-2,0) circle (2.2pt)
node[above=1.5pt] {\color{black} $\sigma_2$};
\filldraw[fill=black,draw=black] (-1,0) circle (2.2pt)
node[above=1.5pt] {\color{black} $\sigma_3$};
\filldraw[fill=black,draw=black] (-3,-1) circle (2.2pt)
node[left=1.5pt] {\color{black} $\sigma_4$};
\filldraw[fill=black,draw=black] (-2,-1) circle (2.2pt)
node[above=1.5pt] {\color{black} $\sigma_5$};
\filldraw[fill=black,draw=black] (-1,-1) circle (2.2pt)
node[right=1.5pt] {\color{black} $\sigma_6$};
\filldraw[fill=black,draw=black] (-3,-2) circle (2.2pt)
node[below=1.5pt] {\color{black} $\sigma_7$};
\filldraw[fill=black,draw=black] (-2,-2) circle (2.2pt)
node[below=1.5pt] {\color{black} $\sigma_8$};
\filldraw[fill=black,draw=black] (-1,-2) circle (2.2pt)
node[below=1.5pt] {\color{black} $\sigma_9$};

\end{tikzpicture}
\caption{Square lattice model where $\sigma_i$ stands for spins and edge interactions are represented with $W(\sigma_i,\sigma_j)$ and $\overline{W}(\sigma_i,\sigma_j)$ for horizontal and vertical, respectively.}
\label{squarelattice}
\end{figure}

As shown in Fig.\ref{squarelattice}, spins are located on sites (vertex) and interactions are driven on edges. Every Boltzmann weight tells the statistics of the interaction between two neighboring spins with the anisotropic interactions. Anisotropy in the lattice distinguishes horizontal and vertical interactions, where the horizontal Boltzmann weight is denoted as $W(\sigma_i,\sigma_j)$ and the vertical one is denoted as $\overline{W}(\sigma_i,\sigma_j)$.

The crucial point is the evaluation of the partition function of the statistical model in the thermodynamic limit where the number of particles in the lattice goes to infinity, $N\to \infty$. If the Boltzmann weights of the model satisfy the star-triangle or the star-star relation, then one can exactly evaluate the partition function.

The square lattice model depicted in Fig.\ref{squarelattice} has the following partition function
 \begin{align}
    Z=\sum \int  \prod_{<i,j>} W(\sigma_i,\sigma_j)\prod_{<k,l>} \overline{W}(\sigma_k,\sigma_l)\prod_{n=1}^N S(\sigma_n)\:dx ,
\end{align}
where $N$ is the total number of spin sites, $<i,j>$ represents the products of only neighbor sites and $S(\sigma_n)$ is for self-interaction at each site.

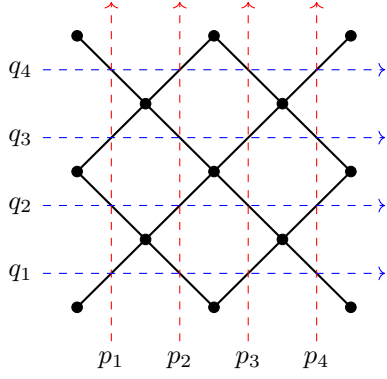
\begin{figure}[tbh]
\centering
\begin{tikzpicture}[scale=0.9]

%\draw[dashed,step=1.0] (0,0) grid  (3.5,3.5);

\draw[-,thick] (-4.5,-0.5)--(-0.5,3.5);
\draw[-,thick] (-4.5,3.5)--(-0.5,-0.5);
\draw[-,thick] (-4.5,1.5)--(-2.5,3.5)--(-0.5,1.5)--(-2.5,-0.5)--(-4.5,1.5);
%\fill[white!] (0,4.5) circle (0.1pt);
\foreach \x in {-4}{
\draw[->,dashed, red] (\x,-1) -- (\x,4);
\fill[white!] (\x,-1) circle (0.1pt)
node[below=0.05pt]{\color{black}\small $p_1$};}
\foreach \x in {-3}{
\draw[->,dashed, red] (\x,-1) -- (\x,4);
\fill[white!] (\x,-1) circle (0.1pt)
node[below=0.05pt]{\color{black}\small $p_2$};}
\foreach \x in {-2}{
\draw[->,dashed, red] (\x,-1) -- (\x,4);
\fill[white!] (\x,-1) circle (0.1pt)
node[below=0.05pt]{\color{black}\small $p_3$};}
\foreach \x in {-1}{
\draw[->,dashed, red] (\x,-1) -- (\x,4);
\fill[white!] (\x,-1) circle (0.1pt)
node[below=0.05pt]{\color{black}\small $p_4$};}
\foreach \y in {0}{
\draw[->,dashed, blue] (-5,\y) -- (0,\y);
\fill[white!] (-5,\y) circle (0.1pt)
node[left=0.05pt]{\color{black}\small $q_1$};}
\foreach \y in {1}{
\draw[->,dashed, blue] (-5,\y) -- (0,\y);
\fill[white!] (-5,\y) circle (0.1pt)
node[left=0.05pt]{\color{black}\small $q_2$};}
\foreach \y in {2}{
\draw[->,dashed, blue] (-5,\y) -- (0,\y);
\fill[white!] (-5,\y) circle (0.1pt)
node[left=0.05pt]{\color{black}\small $q_3$};}
\foreach \y in {3}{
\draw[->,dashed, blue] (-5,\y) -- (0,\y);
\fill[white!] (-5,\y) circle (0.1pt)
node[left=0.05pt]{\color{black}\small $q_4$};}
\foreach \y in {-0.5,1.5,3.5}{
\foreach \x in {-4.5,-2.5,-0.5}{
\filldraw[fill=black,draw=black] (\x,\y) circle (2.2pt);}}
\foreach \y in {0.5,2.5}{
\foreach \x in {-3.5,-1.5}{
\filldraw[fill=black,draw=black] (\x,\y) circle (2.2pt);}}

\end{tikzpicture}
\caption{Rotated square lattice  and rapidity lines.}
\label{rapiditylines}
\end{figure}

In Fig.\ref{rapiditylines}, we draw a new square lattice rotated by $45^\circ$  via introducing horizontal (blue) and vertical (red) rapidity lines since we want to label each interaction with real-valued rapidity parameters $q_i$ and $p_j$ to distinguish each horizontal and each vertical interactions.

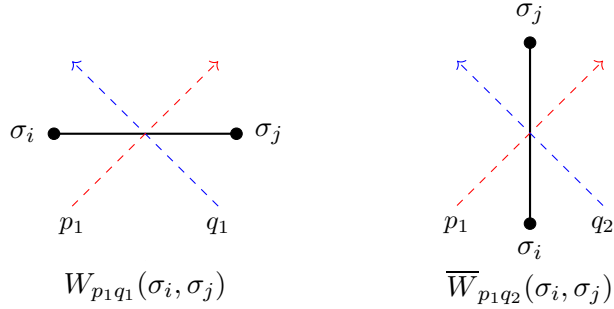
\begin{figure}[tbh]
\centering
\begin{tikzpicture}[scale=2.4]

\draw[-,thick] (-0.5,2)--(0.5,2);
\draw[->,dashed,blue] (0.4,1.6)--(-0.4,2.4);
\fill[white!] (0.4,1.6) circle (0.01pt)
node[below=0.5pt]{\color{black}\small $q_1$};
\draw[->,dashed,red] (-0.4,1.6)--(0.4,2.4);
\fill[white!] (-0.4,1.6) circle (0.01pt)
node[below=0.5pt]{\color{black}\small $p_1$};
\filldraw[fill=black,draw=black] (-0.5,2) circle (0.9pt)
node[left=3pt]{\color{black} $\sigma_i$};
\filldraw[fill=black,draw=black] (0.5,2) circle (0.9pt)
node[right=3pt]{\color{black} $\sigma_j$};

\fill (0,1.3) circle(0.01pt)
node[below=0.05pt]{\color{black} $W_{p_1q_1}(\sigma_i,\sigma_j)$};

\begin{scope}[xshift=60pt,yshift=57pt]
\draw[-,thick] (0,-0.5)--(0,0.5);
\draw[->,dashed,red] (-0.4,-0.4)--(0.4,0.4);
\fill[white!] (-0.4,-0.4) circle (0.01pt)
node[below=0.5pt]{\color{black}\small $p_1$};
\draw[->,dashed,blue] (0.4,-0.4)--(-0.4,0.4);
\fill[white!] (0.4,-0.4) circle (0.01pt)
node[below=0.5pt]{\color{black}\small $q_2$};
\filldraw[fill=black,draw=black] (0,-0.5) circle (0.9pt)
node[below=3pt]{\color{black} $\sigma_i$};
\filldraw[fill=black,draw=black] (0,0.5) circle (0.9pt)
node[above=3pt]{\color{black} $\sigma_j$};

\fill (0,-0.7) circle(0.01pt)
node[below=0.05pt]{\color{black} $\overline{W}_{p_1q_2}(\sigma_i,\sigma_j)$};
\end{scope}
\end{tikzpicture}
\caption{Two types of Boltzmann weight are defined concerning the rapidity lines.}
\label{anisotropicinteractions}
\end{figure}

Then, every spin interaction can be distinguished by the rapidity lines intersecting the edge between the spins, as shown in Figure \ref{anisotropicinteractions}. If directed lines come from the same side before the intersection and leave from the interaction to the same side after the intersection, the Boltzmann weight of the corresponding edge will be denoted as $W_{p_1q_1}(\sigma_i,\sigma_j)$. The other intersection possibility is being oppositely crossed and the interaction at such edges has the Boltzmann weight $\overline{W}_{p_1q_2}(\sigma_i,\sigma_j)$.

In most cases, lattice models have Boltzmann weights as the function of the difference of rapidity parameters. An example of an exception to difference property is the Chiral Potts model \cite{Baxter:1987eq} and it can be obtained by the reduction of "the master" model which has the difference property \cite{Kels:2017vbc}.  

In this paper, both Boltzmann weights $W$ and $\overline{W}$ will be considered as depending only on the difference of the rapidity parameters $q-p$ and, thus, we can rewrite Boltzmann weights depending on the spectral parameter $\alpha$, i.e. the difference of  corresponding rapidity lines $\alpha = p-q$. 

Additionally, let us assume that vertical interaction can be written in the form of horizontal interaction via  crossing parameter. 

\begin{align}
\overline{W}_{\alpha}=W_{\eta-\alpha},
\end{align}
where $\eta$ is positive valued model-dependent parameter and is called the "crossing parameter". 

\begin{figure}[tbh]
\centering
\begin{tikzpicture}[scale=1.25]

%\draw[dashed,step=1.0] (0,0) grid  (3.5,3.5);

\draw[-,thick] (-2.5,1.5)--(-0.5,3.5);
\draw[-,thick] (-4.5,3.5)--(-2.5,1.5);
\draw[-,thick] (-4.5,1.5)--(-2.5,3.5)--(-0.5,1.5);
%\fill[white!] (0,4.5) circle (0.1pt);
\foreach \x in {-4}{
\draw[->,dashed, red] (\x,1) -- (\x,4);
\fill[white!] (\x,1) circle (0.1pt)
node[below=0.05pt]{\color{black}\small $p_1$};}
\foreach \x in {-3}{
\draw[->,dashed, red] (\x,1) -- (\x,4);
\fill[white!] (\x,1) circle (0.1pt)
node[below=0.05pt]{\color{black}\small $p_2$};}
\foreach \x in {-2}{
\draw[->,dashed, red] (\x,1) -- (\x,4);
\fill[white!] (\x,1) circle (0.1pt)
node[below=0.05pt]{\color{black}\small $p_3$};}
\foreach \x in {-1}{
\draw[->,dashed, red] (\x,1) -- (\x,4);
\fill[white!] (\x,1) circle (0.1pt)
node[below=0.05pt]{\color{black}\small $p_4$};}
\foreach \y in {2}{
\draw[->,dashed, blue] (-5,\y) -- (0,\y);
\fill[white!] (-5,\y) circle (0.1pt)
node[left=0.05pt]{\color{black}\small $q_1$};}
\foreach \y in {3}{
\draw[->,dashed, blue] (-5,\y) -- (0,\y);
\fill[white!] (-5,\y) circle (0.1pt)
node[left=0.05pt]{\color{black}\small $q_2$};}
\foreach \y in {1.5,3.5}{
\foreach \x in {-4.5,-2.5,-0.5}{
\filldraw[fill=black,draw=black] (\x,\y) circle (2.2pt);}}
\foreach \y in {2.5}{
\foreach \x in {-3.5,-1.5}{
\filldraw[fill=black,draw=black] (\x,\y) circle (2.2pt);}}

\foreach \x in {-4.5,-2.5,-0.5}{
\fill[white!] (\x,4) circle (0.1pt)
node[below=0.05pt]{\color{black}\small $\sigma$};}
\foreach \x in {-3.5,-1.5}{
\fill[white!] (\x,3) circle (0.1pt)
node[below=0.05pt]{\color{black}\small $\Tilde{\sigma}$};}
\foreach \x in {-4.5,-2.5,-0.5}{
\fill[white!] (\x,2) circle (0.1pt)
node[below=0.05pt]{\color{black}\small $\Tilde{\Tilde{\sigma}}$};}

\foreach \y in {3}{
\fill[white!] (2.5,\y) circle (0.1pt)
node[left=0.05pt]{\color{black}\small $T_{q_2}[\phi(\sigma),\phi(\Tilde{\sigma})]$};}
\foreach \y in {2}{
\fill[white!] (2.5,\y) circle (0.1pt)
node[left=0.05pt]{\color{black}\small $T_{q_1}[\phi(\Tilde{\sigma}),\phi(\Tilde{\Tilde{\sigma}})]$};}

\end{tikzpicture}
\caption{Two transfer matrices $T_{q_1}$ and $T_{q_2}$.}
\label{AB}
\end{figure}
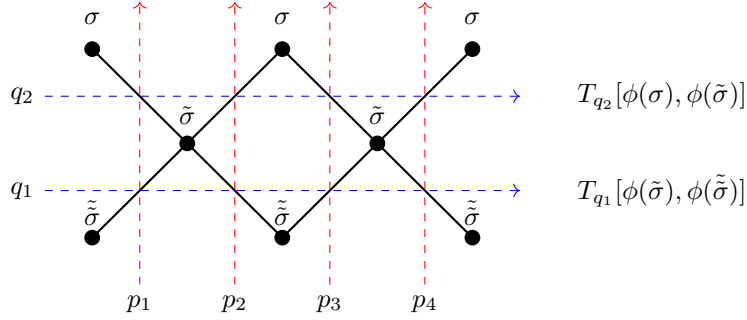

Now, we construct transfer matrices \cite{Baxter:1982zz} and see their commutation relations. As indicated in Fig.\ref{AB}, possible states of every row at the rotated square lattice can be written as an entry of associated transfer matrix written as multiplication of Boltzmann weights at the same row.
        \begin{align}
    T_{q_i}[\phi(\sigma),\phi(\Tilde{\sigma})]=\prod_{j=1}^M W_{p_jq_i}(\sigma_j,\tilde{\sigma}_{j})\overline{W}_{p_jq_i}(\sigma_{j+1},\tilde{\sigma}_{j}),
\end{align}
where $M$ is the number of sites at the row and the function $\phi$ represents all spins at the same row. Thus, we can re-write partition function in terms of transfer matrices 
  \begin{align}
      Z=\sum \int  \prod_{<ij>}
       T_{q_i}[\phi(\sigma),\phi(\Tilde{\sigma})] T_{q_j}[\phi(\Tilde{\sigma}),\phi(\Tilde{\Tilde{\sigma}})] \:.
  \end{align}

The exact evaluation of the partition function is possible if transfer matrices have a commutation relation
\begin{align}
    T_{q_1}[\phi(\Tilde{\sigma}),\phi(\Tilde{\Tilde{\sigma}})]T_{q_2}[\phi(\sigma),\phi(\Tilde{\sigma})]=
T_{q_2}[\phi(\sigma),\phi(\Tilde{\sigma})]T_{q_1}[\phi(\Tilde{\sigma}),\phi(\Tilde{\Tilde{\sigma}})] \:.
\end{align}

\begin{figure}[tbh]
\centering
\begin{tikzpicture}[scale=1.25]

%\draw[dashed,step=1.0] (0,0) grid  (3.5,3.5);

\draw[-,thick] (-2.5,1.5)--(-0.5,3.5);
\draw[-,thick] (-4.5,3.5)--(-2.5,1.5);
\draw[-,thick] (-4.5,1.5)--(-2.5,3.5)--(-0.5,1.5);
%\fill[white!] (0,4.5) circle (0.1pt);
\foreach \x in {-4}{
\draw[->,dashed, red] (\x,1) -- (\x,4);
\fill[white!] (\x,1) circle (0.1pt)
node[below=0.05pt]{\color{black}\small $p_1$};}
\foreach \x in {-3}{
\draw[->,dashed, red] (\x,1) -- (\x,4);
\fill[white!] (\x,1) circle (0.1pt)
node[below=0.05pt]{\color{black}\small $p_2$};}
\foreach \x in {-2}{
\draw[->,dashed, red] (\x,1) -- (\x,4);
\fill[white!] (\x,1) circle (0.1pt)
node[below=0.05pt]{\color{black}\small $p_3$};}
\foreach \x in {-1}{
\draw[->,dashed, red] (\x,1) -- (\x,4);
\fill[white!] (\x,1) circle (0.1pt)
node[below=0.05pt]{\color{black}\small $p_4$};}
\foreach \y in {3}{
\draw[->,dashed, blue] (-5,\y) -- (0,\y);
\fill[white!] (-5,\y) circle (0.1pt)
node[left=0.05pt]{\color{black}\small $q_1$};}
\foreach \y in {2}{
\draw[->,dashed, blue] (-5,\y) -- (0,\y);
\fill[white!] (-5,\y) circle (0.1pt)
node[left=0.05pt]{\color{black}\small $q_2$};}
\foreach \y in {1.5,3.5}{
\foreach \x in {-4.5,-2.5,-0.5}{
\filldraw[fill=black,draw=black] (\x,\y) circle (2.2pt);}}
\foreach \y in {2.5}{
\foreach \x in {-3.5,-1.5}{
\filldraw[fill=black,draw=black] (\x,\y) circle (2.2pt);}}

\foreach \x in {-4.5,-2.5,-0.5}{
\fill[white!] (\x,2) circle (0.1pt)
node[below=0.05pt]{\color{black}\small $\sigma$};}
\foreach \x in {-3.5,-1.5}{
\fill[white!] (\x,3) circle (0.1pt)
node[below=0.05pt]{\color{black}\small $\Tilde{\sigma}$};}
\foreach \x in {-4.5,-2.5,-0.5}{
\fill[white!] (\x,4) circle (0.1pt)
node[below=0.05pt]{\color{black}\small $\Tilde{\Tilde{\sigma}}$};}

\foreach \y in {2}{
\fill[white!] (2.5,\y) circle (0.1pt)
node[left=0.05pt]{\color{black}\small $T_{q_2}[\phi(\sigma),\phi(\Tilde{\sigma})]$};}
\foreach \y in {3}{
\fill[white!] (2.5,\y) circle (0.1pt)
node[left=0.05pt]{\color{black}\small $T_{q_1}[\phi(\Tilde{\sigma}),\phi(\Tilde{\Tilde{\sigma}})]$};}

\end{tikzpicture}
\caption{Commutation figured out as $T_{q_2}T_{q_1}$.}
\label{BA}
\end{figure}
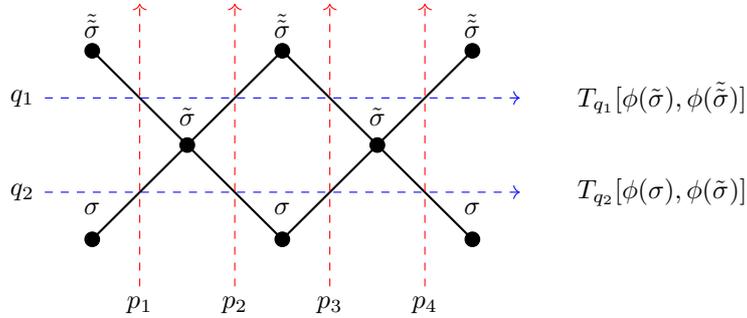

The graphical representation of the commutation of transfer matrices is the equality of Fig.\ref{AB} and Fig.\ref{BA}.

To have a commutation relation, the star-triangle relation or the star-star relations of Boltzmann weights are sufficient. We will consider these sufficient conditions in the following sections.

\subsection{The star-triangle relation and IRF-type Yang-Baxter equation}\label{secstr}

\begin{figure}[tbh]
\centering
\begin{tikzpicture}[scale=2]

\draw[-,thick] (-2,0)--(-2,1);
\draw[-,thick] (-2,0)--(-2.87,-0.5);
\draw[-,thick] (-2,0)--(-1.13,-0.5);
\draw[->,blue,dashed] (-2.9,-0.25)--(-1.1,-0.25);
\fill[white!] (-2.9,-0.25) circle (0.5pt)
node[left=1.5pt]{\color{black}\small $p$};
\draw[->,red,dashed] (-2.7,-0.71)--(-1.7,1.02);
\fill[white!] (-2.7,-0.71) circle (0.5pt)
node[below=1.5pt]{\color{black}\small $q$};
\draw[->,green,dashed] (-1.3,-0.71)--(-2.3,1.02);
\fill[white!] (-1.3,-0.74)circle (0.5pt)
node[below=1.5pt]{\color{black}\small $r$};
\fill (-2,0) circle (1.2pt)
node[below=2.5pt]{\color{black} $\sigma_0$};
\filldraw[fill=black,draw=black] (-2,1) circle (1.2pt)
node[above=1.5pt] {\color{black} $\sigma_i$};
\filldraw[fill=black,draw=black] (-2.87,-0.5) circle (1.2pt)
node[left=1.5pt] {\color{black} $\sigma_k$};
\filldraw[fill=black,draw=black] (-1.13,-0.5) circle (1.2pt)
node[right=1.5pt] {\color{black} $\sigma_j$};

\fill[white!] (0.05,0.3) circle (0.01pt)
node[left=0.05pt] {\color{black}$=$};

\draw[-,thick] (2,1)--(1.13,-0.5);
\draw[-,thick] (1.13,-0.5)--(2.87,-0.5);
\draw[-,thick] (2.87,-0.5)--(2,1);
\draw[->,blue,dashed] (1.1,0.25)--(2.85,0.25);
\fill[white!] (1.1,0.25) circle (0.5pt)
node[left=1.5pt]{\color{black}\small $p $};
\draw[->,red,dashed] (1.75,-0.93)--(2.68,0.67);
\fill[white!] (1.75,-0.93) circle (0.5pt)
node[below=1.5pt]{\color{black}\small $q$};
\draw[->,green,dashed] (2.25,-0.93)--(1.32,0.67);
\fill[white!] (2.25,-0.96) circle (0.5pt)
node[below=1.5pt]{\color{black}\small $r$};
\filldraw[fill=black,draw=black] (2,1) circle (1.2pt)
node[above=1.5pt]{\color{black} $\sigma_i$};
\filldraw[fill=black,draw=black] (1.13,-0.5) circle (1.2pt)
node[left=1.5pt]{\color{black} $\sigma_k$};
\filldraw[fill=black,draw=black] (2.87,-0.5) circle (1.2pt)
node[right=1.5pt]{\color{black} $\sigma_j$};

\end{tikzpicture}
\caption{The star-triangle relation.}
\label{STR}
\end{figure}
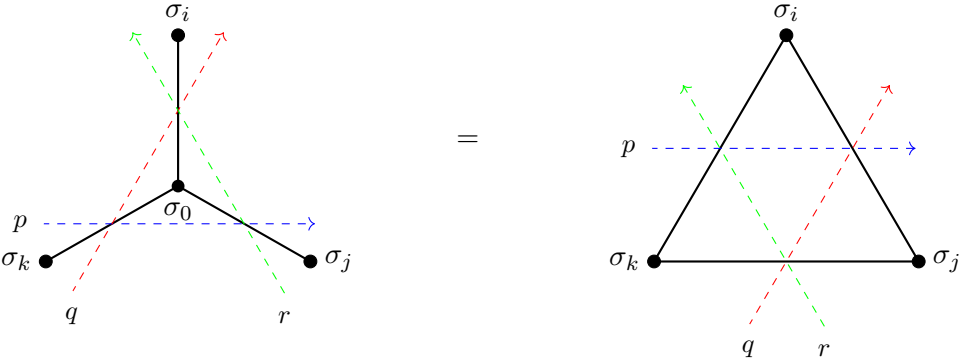

If certain Boltzmann weights $W(\sigma_i, \sigma_j)$ and $\overline{W}(\sigma_i, \sigma_j)$ solve the star-triangle relation, transfer matrices have the commutation property. In the absence of reflection symmetry of spins in the Boltzmann weight $W(\sigma_i, \sigma_j)\neq W(\sigma_j, \sigma_i)$, we need two kinds of star-triangle relation that the relation itself and relation with reflected Boltzmann weights. The following star-triangle equation (depicted in Fig.\ref{STR}) is sufficient as an integrability condition
\begin{align}\nonumber
   \sum_{m_0} \int dx_0 S(\sigma_0) W_{\alpha}(\sigma_1,\sigma_0)W_{\beta}(\sigma_2,\sigma_0)W_{\gamma}(\sigma_3,\sigma_0) 
    \makebox[10em]{}
   \\ \makebox[6em]{}
   =\mathcal{R}(\alpha,\beta,\gamma) W_{\eta-\alpha}(\sigma_1,\sigma_2)W_{\eta-\beta}(\sigma_1,\sigma_3)W_{\eta-\gamma}(\sigma_2,\sigma_3) 
   \label{str},
\end{align}
where $\mathcal{R}(\alpha,\beta,\gamma)$\footnote{In previously known models, there can be also a normalization making that spin-independent function $\mathcal{R}$ equal to one.} stands for a spin-independent function and the crossing parameter becomes $\eta=\alpha+\beta+\gamma$.

Integration and summation in the left-hand side of (\ref{str}) are evaluated over the center spin $\sigma_0=(x_0,m_0)$ as shown in the LHS of Fig.\ref{STR}. 

\begin{figure}[tbh]
\centering
\begin{tikzpicture}[scale=0.9]

\begin{scope}[xshift=-180pt]

% \draw[-,gray] (-2,0)--(-1,-1.73)--(1,-1.73)--(2,0)--(1,1.73)--(-1,1.73)--(-2,0)--(0,0);
% \draw[-,gray] (1,1.73)--(0,0)--(1,-1.73);

\draw[-,very thick] (2,0)--(1,1.73);
\draw[-,very thick] (1,1.73)--(1.70,3.10);
\draw[-,very thick] (1,-1.73)--(2,0);
\draw[-,very thick] (1,-1.73)--(1.70,-3.10);
\draw[-,very thick] (3.30,0)--(2,0);
\draw[-,very thick] (-1,1.73)--(-2,0);
\draw[-,very thick] (-1,-1.73)--(-2,0);
\draw[-,very thick] (-1,-1.73)--(-1.70,-3.10);
\draw[-,very thick] (-1,1.73)--(-1.70,3.10);
\draw[-,very thick] (-3.30,0)--(-2,0);
\draw[-,very thick] (-1,-1.73)--(1,-1.73);
\draw[-,very thick] (-1,1.74)--(1,1.74);

\filldraw[fill=black,draw=black] (-1.7,-3.1) circle (2.0pt)
node[below=1pt]{\small $\sigma_E$};
\filldraw[fill=black,draw=black] (1.7,-3.1) circle (2.0pt)
node[below=1pt]{\small $\sigma_D$};
\filldraw[fill=black,draw=black] (3.3,0) circle (2.0pt)
node[right=1pt]{\small $\sigma_C$};
\filldraw[fill=black,draw=black] (1.7,3.1) circle (2.0pt)
node[above=1pt]{\small $\sigma_B$};
\filldraw[fill=black,draw=black] (-1.7,3.10) circle (2.0pt)
node[above=1pt]{\small $\sigma_A$};
\filldraw[fill=black,draw=black] (-3.1,0) circle (2.0pt)
node[left=1pt]{\small $\sigma_F$};

\filldraw[fill=black,draw=black] (-1,-1.73) circle (2.0pt)
node[below=1pt]{\small };
\filldraw[fill=black,draw=black] (1,-1.73) circle (2.0pt)
node[below=1pt]{\small };
\filldraw[fill=black,draw=black] (2,0) circle (2.0pt)
node[right=1pt]{\small };
\filldraw[fill=black,draw=black] (1,1.73) circle (2.0pt)
node[above=1pt]{\small };
\filldraw[fill=black,draw=black] (-1,1.73) circle (2.0pt)
node[above=1pt]{\small };
\filldraw[fill=black,draw=black] (-2,0) circle (2.0pt)
node[left=1pt]{\small };

\draw[->,dashed,blue]  (-0.5,-2.78)..  controls (-1.5,-2.23) .. (-1.7,-1.73) .. controls (-1,0) .. (-1.7,1.73) .. controls (-1.6,2.33) .. (-0.7,2.71);
\draw[->,dashed,red]  (-2.5,-2.78)..  controls (-0.5,-2.23) .. (0.7,-1.2) .. controls (2.3,-0.8) .. (2.7,1);
\draw[->,dashed,purple]  (-2.7,1)..  controls (-2.3,-0.8) .. (-0.7,-1.2) .. controls (0,-2.23) .. (0.5,-2.33) .. controls (1.2,-2.33) .. (2.7,-2.71);
\draw[->,dashed,green]  (-3.2,-1)..  controls (-2,0.7) .. (-0.9,1) .. controls (0,2.32) .. (2.2,2.32);
\draw[->,dashed,violet]  (0.5,-2.78)..  controls (1.5,-2.23) .. (1.7,-1.73) .. controls (1,0) .. (1.7,1.73) .. controls (1.6,2.33) .. (0.8,2.91);
\draw[->,dashed,yellow]  (-2.8,2.4)..  controls (-1.7,2.3) .. (0.5,1.2) .. controls (2,0.7) .. (2.99,-0.72);

\draw[black!] (-2.95,-1.1) circle (0.01pt)
node[left=1pt]{\color{black}\small $t_1$};
\draw[black!] (-2.45,1.11) circle (0.01pt)
node[left=1pt]{\color{black}\small $t_5$};
\draw[black!] (-2.6,-2.38) circle (0.01pt)
node[below=1pt]{\color{black}\small $t_4$};
\draw[black!] (-0.23,-2.48) circle (0.01pt)
node[below=1pt]{\color{black}\small $t_6$};
\draw[black!] (-2.85,2.45) circle (0.01pt)
node[right=1pt]{\color{black}\small $t_2$};
\draw[black!] (0.20,-2.76) circle (0.01pt)
node[right=1pt]{\color{black}\small $t_3$};

\end{scope}

\end{tikzpicture}
\caption{Lattice form of the composition of Boltzmann weights.}
\label{hexagon}
\end{figure}
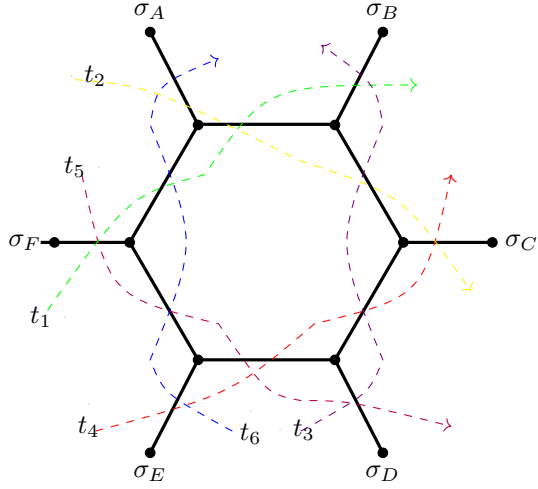

The star-triangle relation allows us also to derive the integrability condition of the IRF-type models. In this derivation, we consider factorized quartic spin interaction in the form of four separate Boltzmann weights coming from the edge interactions. Therefore, we can start with edge interaction lattice in Fig.\ref{hexagon} to obtain the IRF-type Yang-Baxter equation (\ref{IRFstrfac}).

For left hand side of the equation (\ref{IRFstrfac}), we follow these steps on Fig.\ref{hexagon}. 

\begin{itemize}
    \item Apply star-triangle relation on odd sites $\sigma_1$, $\sigma_3$, and $\sigma_5$ to go from star to triangle.
    \item Use one more time the same relation converting triangle to star, that is, obtain star with the center spin $\sigma_0$.
\end{itemize}
When we follow the same steps but for different sites, we acquire right hand side of the equation (\ref{IRFstrfac}).
\begin{itemize}
    \item Apply star-triangle relation on odd sites $\sigma_2$, $\sigma_4$ and $\sigma_6$ to go from star to triangle.
    \item Once again use the same relation to convert triangle to star with the center spin $\sigma_0$.
\end{itemize}

An important point is that rapidity lines should be directed carefully after all transformations.

\begin{figure}[tbh]
\centering
\begin{tikzpicture}[scale=1.3]

\draw[-,very thick] (9,0)--(7,0);\draw[-,very thick] (8,1.73)--(8,-1.73);
\draw[-,very thick] (8,1.73)--(5,0);\draw[-,very thick] (6,1.73)--(7,0);
\draw[-,very thick] (8,-1.73)--(5,0);\draw[-,very thick] (6,-1.73)--(7,0);

\filldraw[fill=black,draw=black] (6,-1.73) circle (2.0pt)
node[below=1pt]{\small $\sigma_E$};
\filldraw[fill=black,draw=black] (8,-1.73) circle (2.0pt)
node[below=1pt]{\small $\sigma_D$};
\filldraw[fill=black,draw=black] (9,0) circle (2.0pt)
node[right=1pt]{\small $\sigma_C$};
\filldraw[fill=black,draw=black] (8,1.73) circle (2.0pt)
node[above=1pt]{\small $\sigma_B$};
\filldraw[fill=black,draw=black] (6,1.73) circle (2.0pt)
node[above=1pt]{\small $\sigma_A$};
\filldraw[fill=black,draw=black] (5,0) circle (2.0pt)
node[left=1pt]{\small $\sigma_F$};
\filldraw[fill=black,draw=black] (7,0) circle (2.0pt)
node[below=4pt]{\small $\sigma_0$};
\filldraw[fill=black,draw=black] (6.5,0.87) circle (2.0pt);
\filldraw[fill=black,draw=black] (6.5,-0.87) circle (2.0pt);
\filldraw[fill=black,draw=black] (8,0) circle (2.0pt);

\draw[->,dashed,blue]  (-0.3,-1.88)..  controls (-0.2,-1.73) .. (-0.1,-0.73) .. controls (0.6,0) .. (0.1,0.53) .. controls (-0.4,1.33) .. (-0,1.91);
\draw[->,dashed,red]  (-2.2,-1.48)..  controls (-0.5,-0.5) .. (-0.2,0.4) .. controls (0.3,0.3) .. (2,0.9);
\draw[->,dashed,purple]  (-1.9,0.7)..  controls (-1.5,0) .. (-1,-0.78) .. controls (-0.2,-1.23) .. (0.5,-1.33) .. controls (1,-1.3) .. (1.7,-1.4);
\draw[->,dashed,green]  (-1.7,-0.7)..  controls (-1.4,0) .. (-1,1) .. controls (-0.3,1.43) .. (0.5,1.32) .. controls (1,1.3) .. (1.7,1.41);
\draw[->,dashed,pink]  (0.3,-1.9)..  controls (1.2,-1) .. (1.1,-0.53) .. controls (1.3,0) .. (1.1,0.53) .. controls (1.2,1) .. (0.4,1.8);
\draw[->,dashed,violet]  (-2.2,1.4)..  controls (-0.7,0.5) .. (-0.2,-0.5) .. controls (0.5,-0.2) .. (2,-0.82);

\draw[black!] (-1.5,-0.84) circle (0.01pt)
node[left=1pt]{\color{black}\small $t_1$};
\draw[black!] (-1.45,0.81) circle (0.01pt)
node[left=1pt]{\color{black}\small $t_5$};
\draw[black!] (-2.3,1.68) circle (0.01pt)
node[below=1pt]{\color{black}\small $t_2$};
\draw[black!] (0.3,-1.88) circle (0.01pt)
node[below=1pt]{\color{black}\small $t_3$};
\draw[black!] (-2.5,-1.51) circle (0.01pt)
node[right=1pt]{\color{black}\small $t_4$};
\draw[black!] (-0.45,-1.96) circle (0.01pt)
node[right=1pt]{\color{black}\small $t_6$};

\draw[-,very thick] (-2,0)--(0,0);
\draw[-,very thick] (-1,1.73)--(-1,-1.73);
\draw[-,very thick] (-1,1.73)--(2,0);
\draw[-,very thick] (1,1.73)--(0,0);
\draw[-,very thick] (-1,-1.73)--(2,0);
\draw[-,very thick] (1,-1.73)--(0,0);

\filldraw[fill=black,draw=black] (-1,-1.73) circle (2.0pt)
node[below=1pt]{\small $\sigma_E$};
\filldraw[fill=black,draw=black] (1,-1.73) circle (2.0pt)
node[below=1pt]{\small $\sigma_D$};
\filldraw[fill=black,draw=black] (2,0) circle (2.0pt)
node[right=1pt]{\small $\sigma_C$};
\filldraw[fill=black,draw=black] (1,1.73) circle (2.0pt)
node[above=1pt]{\small $\sigma_B$};
\filldraw[fill=black,draw=black] (-1,1.73) circle (2.0pt)
node[above=1pt]{\small $\sigma_A$};
\filldraw[fill=black,draw=black] (-2,0) circle (2.0pt)
node[left=1pt]{\small $\sigma_F$};
\filldraw[fill=black,draw=black] (0,0) circle (2.0pt)
node[below=4pt]{\small $\sigma_0$};
\filldraw[fill=black,draw=black] (0.5,0.87) circle (2.0pt);
\filldraw[fill=black,draw=black] (0.5,-0.87) circle (2.0pt);
\filldraw[fill=black,draw=black] (-1,0) circle (2.0pt);

\draw[->,dashed,blue]  (6.5,-1.78)..  controls (6,-1.23) .. (5.8,-0.3) .. controls (5.73,0) .. (5.8,0.3) .. controls (6,1.33) .. (6.5,1.71);
\draw[->,dashed,red]  (5,-1.2)..  controls (5.9,-0.43) .. (7.2,-0.3) .. controls (7.5,0.2) .. (8.7,1.3);
\draw[->,dashed,purple]  (5,1)..  controls (5.7,0.5) .. (6.7,0.4) .. controls (7.2,0.23) .. (7.5,0) .. controls (8,-0.8) .. (8.7,-1.71);
\draw[->,dashed,green]  (5.6,-1.55)..  controls (7,-1.3) .. (8,-0.7) .. controls (8.3,0) .. (9,0.52);
\draw[->,dashed,pink]  (7.2,-1.78)..  controls (7,-1.23) .. (6.7,-0.53) .. controls (6.5,0) .. (6.7,0.53) .. controls (7,1.33) .. (7.2,1.91);
\draw[->,dashed,violet]  (5.4,1.5)..  controls (7,1.3) .. (8,0.7) .. controls (8.3,0) .. (9,-0.52);

\draw[black!] (7.1,-1.84) circle (0.01pt)
node[right=1pt]{\color{black}\small $t_3$};
\draw[black!] (6.5,-1.91) circle (0.01pt)
node[right=1pt]{\color{black}\small $t_6$};
\draw[black!] (5.4,-1.4) circle (0.01pt)
node[below=1pt]{\color{black}\small $t_4$};
\draw[black!] (4.9,-1) circle (0.01pt)
node[below=1pt]{\color{black}\small $t_1$};
\draw[black!] (5,1) circle (0.01pt)
node[left=1pt]{\color{black}\small $t_5$};
\draw[black!] (5.4,1.6) circle (0.01pt)
node[left=1pt]{\color{black}\small $t_2$};

\draw[black] (3.9,0) circle (0.01pt)
node[left=1pt]{\color{black}\small $=$};

\end{tikzpicture}
\caption{ IRF-type Yang-Baxter relation obtained by operating the star-triangle relation over Fig.\ref{hexagon}.}
\label{STRYBE}
\end{figure}
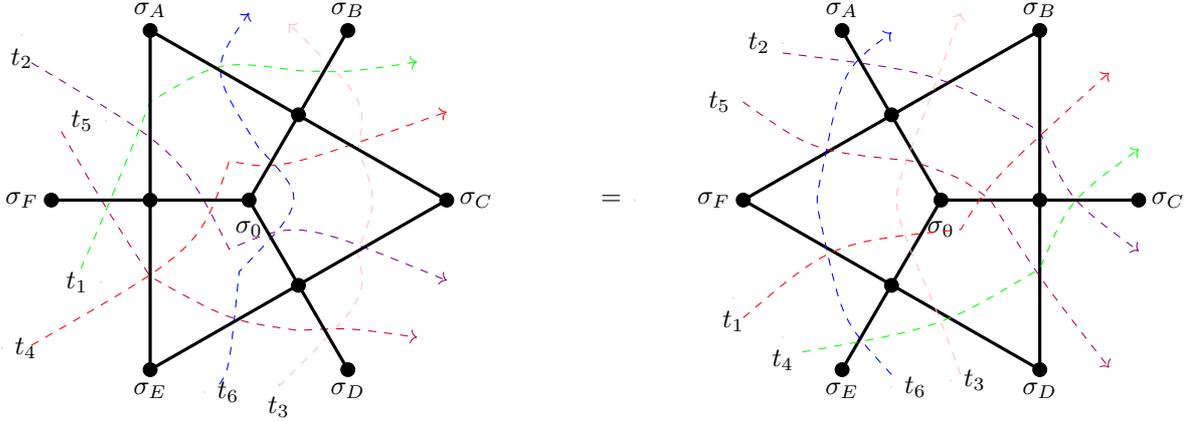
We obtained pictorial equality in Fig.\ref{STRYBE} and to write it as YBE one needs to define factorized Boltzmann weight over the face

\begin{align}
R_{t_{34} t_{21}}\left(\begin{array}{cc}
\sigma_{4} & \sigma_{3} \\
\sigma_{1} & \sigma_{2}
\end{array}\right)=\sum_{m_{i}} \int d x_{i} W_{t_{32}}\left(\sigma_{1}, \sigma_{i}\right) W_{t_{24}}\left(\sigma_{i}, \sigma_{2}\right) W_{t_{41}}\left(\sigma_{3}, \sigma_{i}\right) W_{t_{13}}\left(\sigma_{i}, \sigma_{4}\right) \:.
\label{factorized}
\end{align}

By the definition of Boltzmann weight of interaction round a unit face, the Yang-Baxter equation can be written in the IRF-type

\begin{align}
    \begin{aligned}
\sum_{m_0} \int dx_0  \; R_{t_{25}t_{41}}\left( \begin{array}{cc}
  {\sigma_A} & {\sigma_0}\\ {\sigma_F} & 
  {\sigma_E}\end{array}\right) \; R_{t_{63}t_{25}}\left(\begin{array}{cc}
    {\sigma_0} & {\sigma_C}\\ {\sigma_E} &
    {\sigma_D}\end{array}\right)
 R_{t_{41}t_{63}}\left(\begin{array}{cc}
      {\sigma_A} & {\sigma_B}\\ {\sigma_0} &
      {\sigma_C}\end{array}\right)  \makebox[4em]{}
      \\
\makebox[4em]{} 
= \sum_{m_0 } \int dx_0 \; R_{t_{41}t_{63}}\left(\begin{array}{cc}
        {\sigma_F} & {\sigma_0}\\ {\sigma_E} &
        {\sigma_D}\end{array}\right)
 R_{t_{63}t_{25}}\left(\begin{array}{cc}
          {\sigma_A} & {\sigma_B}\\ {\sigma_F} &
          {\sigma_0}\end{array}\right) \; R_{t_{25}t_{41}}\left(\begin{array}{cc}
            {\sigma_B} & {\sigma_C}\\ {\sigma_0} &
            {\sigma_D}\end{array}\right) \:, \label{IRFstrfac}
\end{aligned}
\end{align}

where $t_{ij}$ represents intersections of two rapidity lines in the same IRF-type Boltzmann weight. One can figure the interaction round a face and it has the same construction \cite{perk1986nonintersecting} as the star-triangle relation in edge interaction.

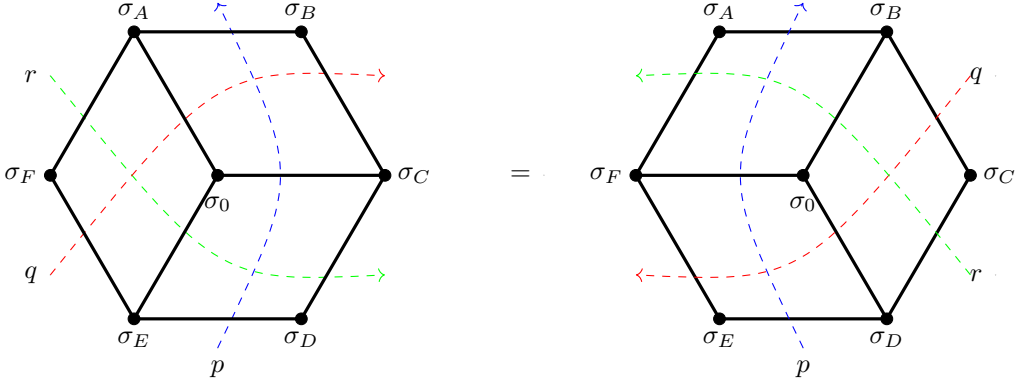
\begin{figure}[tbh!]
\centering
\begin{tikzpicture}[scale=1.1]

\draw[-,very thick] (5,0)--(7,0);
\draw[-,very thick] (5,0)--(6,-1.73);
\draw[-,very thick] (5,0)--(6,1.73);
\draw[-,very thick] (6,1.73)--(8,1.73);
\draw[-,very thick] (6,-1.73)--(8,-1.73);
\draw[-,very thick] (8,1.73)--(7,0);
\draw[-,very thick] (8,-1.73)--(7,0);
\draw[-,very thick] (8,-1.73)--(9,0);
\draw[-,very thick] (8,1.73)--(9,0);

\filldraw[fill=black,draw=black] (6,-1.73) circle (2.0pt)
node[below=1pt]{\small $\sigma_E$};
\filldraw[fill=black,draw=black] (8,-1.73) circle (2.0pt)
node[below=1pt]{\small $\sigma_D$};
\filldraw[fill=black,draw=black] (9,0) circle (2.0pt)
node[right=1pt]{\small $\sigma_C$};
\filldraw[fill=black,draw=black] (8,1.73) circle (2.0pt)
node[above=1pt]{\small $\sigma_B$};
\filldraw[fill=black,draw=black] (6,1.73) circle (2.0pt)
node[above=1pt]{\small $\sigma_A$};
\filldraw[fill=black,draw=black] (5,0) circle (2.0pt)
node[left=1pt]{\small $\sigma_F$};
\filldraw[fill=black,draw=black] (7,0) circle (2.0pt)
node[below=4pt]{\small $\sigma_0$};

\draw[black!] (9.3,-1.2) circle (0.01pt)
node[left=1pt]{\color{black}\small $r$};
\draw[black!] (9.3,1.2) circle (0.01pt)
node[left=1pt]{\color{black}\small $q$};
\draw[black!] (7,-2.08) circle (0.01pt)
node[below=1pt]{\color{black}\small $p$};

\draw[->,dashed,blue]  (7,-2.08)..  controls (6,0) .. (7,2.08) ;
\draw[->,dashed,red]  (9,1.2)..  controls (7,-1.3) .. (5,-1.2) ;
\draw[->,dashed,green]  (9,-1.2)..  controls (7,1.3) .. (5,1.2) ;

\draw[black!] (-2,-1.2) circle (0.01pt)
node[left=1pt]{\color{black}\small $q$};
\draw[black!] (-2,1.2) circle (0.01pt)
node[left=1pt]{\color{black}\small $r$};
\draw[black!] (0,-2.08) circle (0.01pt)
node[below=1pt]{\color{black}\small $p$};

\draw[->,dashed,green]  (-2,1.2)..  controls (0,-1.3) .. (2,-1.2);
\draw[->,dashed,red]  (-2,-1.2)..  controls (0,1.3) .. (2,1.2) ;
\draw[->,dashed,blue]  (0,-2.08)..  controls (1,0) .. (0,2.08);

\draw[-,very thick] (0,0)--(2,0);
\draw[-,very thick] (-2,0)--(-1,1.73);
\draw[-,very thick] (-2,0)--(-1,-1.73);
\draw[-,very thick] (2,0)--(1,1.73);
\draw[-,very thick] (0,0)--(-1,-1.73);
\draw[-,very thick] (0,0)--(-1,1.73);
\draw[-,very thick] (2,0)--(1,-1.73);
\draw[-,very thick] (-1,1.73)--(1,1.73);
\draw[-,very thick] (-1,-1.73)--(1,-1.73);

\filldraw[fill=black,draw=black] (-1,-1.73) circle (2.0pt)
node[below=1pt]{\small $\sigma_E$};
\filldraw[fill=black,draw=black] (1,-1.73) circle (2.0pt)
node[below=1pt]{\small $\sigma_D$};
\filldraw[fill=black,draw=black] (2,0) circle (2.0pt)
node[right=1pt]{\small $\sigma_C$};
\filldraw[fill=black,draw=black] (1,1.73) circle (2.0pt)
node[above=1pt]{\small $\sigma_B$};
\filldraw[fill=black,draw=black] (-1,1.73) circle (2.0pt)
node[above=1pt]{\small $\sigma_A$};
\filldraw[fill=black,draw=black] (-2,0) circle (2.0pt)
node[left=1pt]{\small $\sigma_F$};
\filldraw[fill=black,draw=black] (0,0) circle (2.0pt)
node[below=4pt]{\small $\sigma_0$};

\draw[black] (3.9,0) circle (0.01pt)
node[left=1pt]{\color{black}\small $=$};

\end{tikzpicture}
\caption{IRF-type star-triangle relation with spectral parameters.}
\label{strybeirf}
\end{figure}

Mathematical equality of Fig.\ref{strybeirf}, that is, IRF-type YBE is the following equation

   \begin{align}
    \begin{aligned}
\sum_{m_0 } \int dx_0  \; R_{{qr}}\left( \begin{array}{cc}
  {\sigma_A} & {\sigma_0}\\ {\sigma_F} & 
  {\sigma_E}\end{array}\right) \; R_{{pr}}\left(\begin{array}{cc}
    {\sigma_0} & {\sigma_C}\\ {\sigma_E} &
    {\sigma_D}\end{array}\right)
 R_{{pq}}\left(\begin{array}{cc}
      {\sigma_A} & {\sigma_B}\\ {\sigma_0} &
      {\sigma_C}\end{array}\right)  \makebox[4em]{}
      \\
\makebox[4em]{} 
= \sum_{m_0 } \int dx_0 \; R_{{pq}}\left(\begin{array}{cc}
        {\sigma_F} & {\sigma_0}\\ {\sigma_E} &
        {\sigma_D}\end{array}\right)
 R_{{pr}}\left(\begin{array}{cc}
          {\sigma_A} & {\sigma_B}\\ {\sigma_F} &
          {\sigma_0}\end{array}\right) \; R_{{qr}}\left(\begin{array}{cc}
            {\sigma_B} & {\sigma_C}\\ {\sigma_0} &
            {\sigma_D}\end{array}\right) \:.\label{IRFstr}
\end{aligned}
\end{align}

It should also be noted that the integration and summation in (\ref{IRFstr}) over center spins appear on both sides of the equation unlike the star-triangle relation in Ising-like models. 

\subsection{The star-star relation and IRF-type Yang-Baxter equation}
\begin{figure}[tbh]
\centering
\begin{tikzpicture}[scale=1]

\draw[-,thick] (1,0)--(3,0);
\draw[-,black,thick] (3,2)--(3,0);
\draw[-,black,thick] (5,0)--(3,2);
\draw[-,black,thick] (3,0)--(5,0);
\draw[-,black,thick] (5,0)--(3,-2);
\draw[-,black,thick] (3,0)--(3,-2);

\draw[black!] (2.1,-1.1) circle (0.01pt)
node[left=1pt]{\color{black}\small $t_1$};
\draw[black!] (1.75,-0.58) circle (0.01pt)
node[left=1pt]{\color{black}\small $t_2$};
\draw[black!] (1.5,0.88) circle (0.01pt)
node[below=1pt]{\color{black}\small $t_3$};
\draw[black!] (1.8,1.4) circle (0.01pt)
node[below=1pt]{\color{black}\small $t_4$};

\draw[->,dashed,blue]  (1.7,-0.58)..  controls (2,0) .. (2.4,0.5) .. controls (3.5,1) .. (5,0.8);
\draw[->,dashed,green]  (1.7,0.58)..  controls (2,0) .. (2.4,-0.5) .. controls (3.5,-1) .. (5,-0.8);
\draw[->,dashed,red]  (2,1)..  controls (3,0.7) .. (3.8,0) .. controls (4,-0.83) .. (5,-1.3);
\draw[->,dashed,violet]  (2,-1)..  controls (3,-0.7) .. (3.8,0) .. controls (4,0.83) .. (5,1.32);

\filldraw[fill=black,draw=black] (1,0) circle (1.2pt)
node[left=1.5pt] {\color{black} $\sigma_1$};
\filldraw[fill=black,draw=black] (5,0) circle (1.2pt)
node[right=1.5pt] {\color{black} $\sigma_3$};
\filldraw[fill=black,draw=black] (3,2) circle (1.2pt)
node[above=1.5pt] {\color{black} $\sigma_2$};
\filldraw[fill=black,draw=black] (3,-2) circle (1.2pt)
node[right=1.5pt] {\color{black} $\sigma_4$};
\filldraw[fill=white,draw=black] (3,0) circle (1.2pt)
node[above=1.5pt] {\color{black} $\sigma_A$};

\fill[white!] (6.75,0) circle (0.01pt)
node[left=0.05pt] {\color{black}$=$};

\draw[-,black,thick] (8,0)--(10,0);
\draw[-,thick] (10,2)--(10,0);
\draw[-,thick] (8,0)--(10,-2);
\draw[-,black,thick] (10,0)--(12,0);
\draw[-,thick] (8,0)--(10,2);
\draw[-,thick] (10,0)--(10,-2);

\draw[black!] (11.6,-1.1) circle (0.01pt)
node[left=1pt]{\color{black}\small $t_1$};
\draw[black!] (11.9,-0.68) circle (0.01pt)
node[left=1pt]{\color{black}\small $t_2$};
\draw[black!] (11.5,0.88) circle (0.01pt)
node[below=1pt]{\color{black}\small $t_3$};
\draw[black!] (11.2,1.4) circle (0.01pt)
node[below=1pt]{\color{black}\small $t_4$};

\draw[->,dashed,blue]  (11.3,-0.58)..  controls (11,0) .. (10.6,0.5) .. controls (9.5,1) .. (8,0.8);
\draw[->,dashed,green]  (11.3,0.58)..  controls (11,0) .. (10.6,-0.5) .. controls (9.5,-1) .. (8,-0.8);
\draw[->,dashed,red]  (11,1)..  controls (10,0.7) .. (9.2,0) .. controls (9,-0.83) .. (8,-1.3);
\draw[->,dashed,violet]  (11,-1)..  controls (10,-0.7) .. (9.2,0) .. controls (9,0.83) .. (8,1.32);

\filldraw[fill=black,draw=black] (8,0) circle (1.2pt)
node[left=1.5pt] {\color{black} $\sigma_1$};
\filldraw[fill=black,draw=black] (12,0) circle (1.2pt)
node[right=1.5pt] {\color{black} $\sigma_3$};
\filldraw[fill=black,draw=black] (10,2) circle (1.2pt)
node[above=1.5pt] {\color{black} $\sigma_2$};
\filldraw[fill=black,draw=black] (10,-2) circle (1.2pt)
node[right=1.5pt] {\color{black} $\sigma_4$};
\filldraw[fill=white,draw=black] (10,0) circle (1.2pt)
node[above=1.5pt] {\color{black} $\sigma_A$};

\end{tikzpicture}
\label{ssr}
\caption{The star-star relation.}
\end{figure}
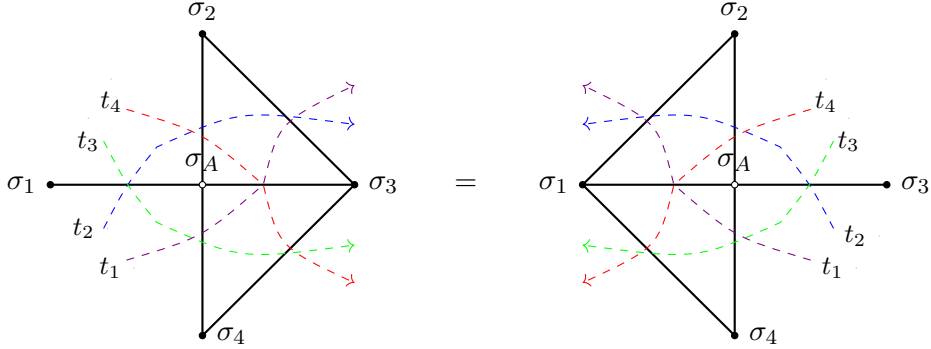

One can easily derive another sufficient condition the star-star relation for the integrability of the statistical lattice spin model by using the star-triangle relation\footnote{There is an example of the three-layer Zamolodchikov model \cite{baxter:1997ssr} which has star-star relation without the star-triangle relation. }.   In the existence of the star-triangle relation, the equality represented in Fig.\ref{ssr} can be seen by the star-triangle relations in four steps from left to right or vice versa.

One ensures that the model is integrable if Boltzmann weights hold the relation below
\begin{align}
    R_{t_{34}t_{21}}\left(\begin{array}{ccc}
       & \sigma_1 &\\
\sigma_2         & & \sigma_3  \\
&\sigma_4  &
    \end{array}\right)=R_{t_{21}t_{34}}\left(\begin{array}{ccc}
       & \sigma_1 &\\
\sigma_2         & & \sigma_3  \\
&\sigma_4  &
    \end{array}\right) \:,\label{sseq}
\end{align}
where 
\begin{align}
    \textcolor{red}{R_{t_{34}t_{21}}\left(\begin{array}{ccc}
       & \sigma_1 &\\
\sigma_2         & & \sigma_3  \\
&\sigma_4  &
    \end{array}\right)}=W_{t_{21}}(\sigma_2,\sigma_3)W_{t_{34}}(\sigma_3,\sigma_4)R_{t_{34}t_{21}}\left(\begin{array}{cc}
    \sigma_4     & \sigma_3 \\
\sigma_1         & \sigma_2
    \end{array}\right) \;.
\end{align}

We only introduce rapidity lines and the spectral parameter and related conditions are discussed in detail \cite{baxter:1997ssr}.

We can re-write the star-star equation with the previous definitions
\begin{align}
R_{t_{34}t_{21}}\left(\begin{array}{cc}
    \sigma_4     & \sigma_3 \\
\sigma_1         & \sigma_2
    \end{array}\right)
=\frac{W_{t_{21}}(\sigma_1,\sigma_2)W_{t_{34}}(\sigma_1,\sigma_4)}{W_{t_{21}}(\sigma_2,\sigma_3)W_{t_{34}}(\sigma_3,\sigma_4)}
R_{t_{21}t_{34}}\left(\begin{array}{cc}
    \sigma_4    & \sigma_3 \\
\sigma_1         & \sigma_2
    \end{array}\right) \:.
    \label{gf}
\end{align}

\begin{figure}[tbh]
\centering
\begin{tikzpicture}[scale=1.1]

% \draw[-,gray] (-2,0)--(-1,-1.73)--(1,-1.73)--(2,0)--(1,1.73)--(-1,1.73)--(-2,0)--(0,0);
% \draw[-,gray] (1,1.73)--(0,0)--(1,-1.73);

\draw[-,very thick,red] (9,0)--(7,0);
\draw[-,very thick,red] (8,1.73)--(8,-1.73);
\draw[-,very thick,blue] (8,1.73)--(5,0);
\draw[-,very thick,blue] (6,1.73)--(7,0);
\draw[-,very thick,green] (8,-1.73)--(5,0);
\draw[-,very thick,green] (6,-1.73)--(7,0);

\draw[-,very thick,red] (8,1.73)--(9,0);
\draw[-,very thick,red] (8,-1.73)--(7,0);

\filldraw[fill=black,draw=black] (6,-1.73) circle (2.0pt)
node[below=1pt]{\small $\sigma_E$};
\filldraw[fill=black,draw=black] (8,-1.73) circle (2.0pt)
node[below=1pt]{\small $\sigma_D$};
\filldraw[fill=black,draw=black] (9,0) circle (2.0pt)
node[right=1pt]{\small $\sigma_C$};
\filldraw[fill=black,draw=black] (8,1.73) circle (2.0pt)
node[above=1pt]{\small $\sigma_B$};
\filldraw[fill=black,draw=black] (6,1.73) circle (2.0pt)
node[above=1pt]{\small $\sigma_A$};
\filldraw[fill=black,draw=black] (5,0) circle (2.0pt)
node[left=1pt]{\small $\sigma_F$};
\filldraw[fill=black,draw=black] (7,0) circle (2.0pt)
node[below=4pt]{\small $\sigma_0$};
\filldraw[fill=black,draw=black] (6.5,0.87) circle (2.0pt);
\filldraw[fill=black,draw=black] (6.5,-0.87) circle (2.0pt);
\filldraw[fill=black,draw=black] (8,0) circle (2.0pt);

\draw[->,dashed,blue]  (-0.3,-1.88)..  controls (-0.2,-1.73) .. (-0.1,-0.73) .. controls (0.6,0) .. (0.1,0.53) .. controls (-0,1.33) .. (-0,1.91);
\draw[->,dashed,red]  (-2,-0.8)..  controls (-0.5,-0.4) .. (-0.4,0.8) .. controls (0.3,1.33) .. (2,1.1);
\draw[->,dashed,purple]  (-1.9,0.7)..  controls (-1.5,0) .. (-1,-0.78) .. controls (-0.2,-1.23) .. (0.5,-1.33) .. controls (1,-1.3) .. (1.7,-1.4);
\draw[->,dashed,green]  (-1.7,-1.4)..  controls (-1.4,0) .. (-1,0.7) .. controls (0,0.5) .. (0.3,0.3) .. controls (1,0.4) .. (1.7,0.61);
\draw[->,dashed,pink]  (0.3,-1.9)..  controls (1.2,-1) .. (1.1,-0.53) .. controls (1.3,0) .. (1.1,0.53) .. controls (1.2,1) .. (0.4,1.8);
\draw[->,dashed,violet]  (-2.2,1.4)..  controls (-0.7,0.5) .. (-0.2,-0.5) .. controls (0.5,-0.2) .. (2,-0.82);

\draw[black!] (-1.5,-1.64) circle (0.01pt)
node[left=1pt]{\color{black}\small $t_1$};
\draw[black!] (-1.45,0.81) circle (0.01pt)
node[left=1pt]{\color{black}\small $t_5$};
\draw[black!] (-2.3,1.68) circle (0.01pt)
node[below=1pt]{\color{black}\small $t_2$};
\draw[black!] (0.3,-1.88) circle (0.01pt)
node[below=1pt]{\color{black}\small $t_3$};
\draw[black!] (-2.3,-0.8) circle (0.01pt)
node[right=1pt]{\color{black}\small $t_4$};
\draw[black!] (-0.45,-1.96) circle (0.01pt)
node[right=1pt]{\color{black}\small $t_6$};

\draw[-,very thick,red] (-2,0)--(0,0);
\draw[-,very thick,red] (-1,1.73)--(-1,-1.73);
\draw[-,very thick,blue] (-1,1.73)--(2,0);
\draw[-,very thick,blue] (1,1.73)--(0,0);
\draw[-,very thick,green] (-1,-1.73)--(2,0);
\draw[-,very thick,green] (1,-1.73)--(0,0);

\draw[-,very thick,red] (-2,0)--(-1,-1.73);
\draw[-,very thick,red] (-1,1.73)--(0,0);

\filldraw[fill=black,draw=black] (-1,-1.73) circle (2.0pt)
node[below=1pt]{\small $\sigma_E$};
\filldraw[fill=black,draw=black] (1,-1.73) circle (2.0pt)
node[below=1pt]{\small $\sigma_D$};
\filldraw[fill=black,draw=black] (2,0) circle (2.0pt)
node[right=1pt]{\small $\sigma_C$};
\filldraw[fill=black,draw=black] (1,1.73) circle (2.0pt)
node[above=1pt]{\small $\sigma_B$};
\filldraw[fill=black,draw=black] (-1,1.73) circle (2.0pt)
node[above=1pt]{\small $\sigma_A$};
\filldraw[fill=black,draw=black] (-2,0) circle (2.0pt)
node[left=1pt]{\small $\sigma_F$};
\filldraw[fill=black,draw=black] (0,0) circle (2.0pt)
node[below=4pt]{\small $\sigma_0$};
\filldraw[fill=black,draw=black] (0.5,0.87) circle (2.0pt);
\filldraw[fill=black,draw=black] (0.5,-0.87) circle (2.0pt);
\filldraw[fill=black,draw=black] (-1,0) circle (2.0pt);

\draw[->,dashed,blue]  (6.5,-1.78)..  controls (6,-1.23) .. (5.8,-0.3) .. controls (5.73,0) .. (5.8,0.3) .. controls (6,1.33) .. (6.5,1.71);
\draw[->,dashed,red]  (5,-1.2)..  controls (5.9,-0.43) .. (7.4,-0.7) .. controls (7.8,-0.6) .. (8,-0.6) .. controls (8.2,-0.2) .. (8.5,0.8) .. controls (9,1.2) .. (9.2,1.5);
\draw[->,dashed,purple]  (5,1)..  controls (5.7,0.5) .. (6.7,0.4) .. controls (7.2,0.23) .. (7.5,0) .. controls (8,-0.8) .. (8.7,-1.71);
\draw[->,dashed,green]  (5.6,-1.55)..  controls (7,-1.3) .. (8,0.8) .. controls (8.3,0.9) .. (9.3,0.92);
\draw[->,dashed,pink]  (7.2,-1.78)..  controls (7,-1.23) .. (6.7,-0.53) .. controls (6.5,0) .. (6.7,0.53) .. controls (7,1.33) .. (7.2,1.91);
\draw[->,dashed,violet]  (5.4,1.5)..  controls (7,1.3) .. (8,0.7) .. controls (8.3,0) .. (8.6,-0.52);

\draw[black!] (7.1,-1.84) circle (0.01pt)
node[right=1pt]{\color{black}\small $t_3$};
\draw[black!] (6.5,-1.91) circle (0.01pt)
node[right=1pt]{\color{black}\small $t_6$};
\draw[black!] (5.4,-1.4) circle (0.01pt)
node[below=1pt]{\color{black}\small $t_4$};
\draw[black!] (4.9,-1) circle (0.01pt)
node[below=1pt]{\color{black}\small $t_1$};
\draw[black!] (5,1) circle (0.01pt)
node[left=1pt]{\color{black}\small $t_5$};
\draw[black!] (5.4,1.6) circle (0.01pt)
node[left=1pt]{\color{black}\small $t_2$};

\draw[black] (3.9,0) circle (0.01pt)
node[left=1pt]{\color{black}\small $=$};

\end{tikzpicture}
\caption{IRF-type Yang-Baxter relation derived by using the star-star relation.}
\label{SSRYBE1}
\end{figure}
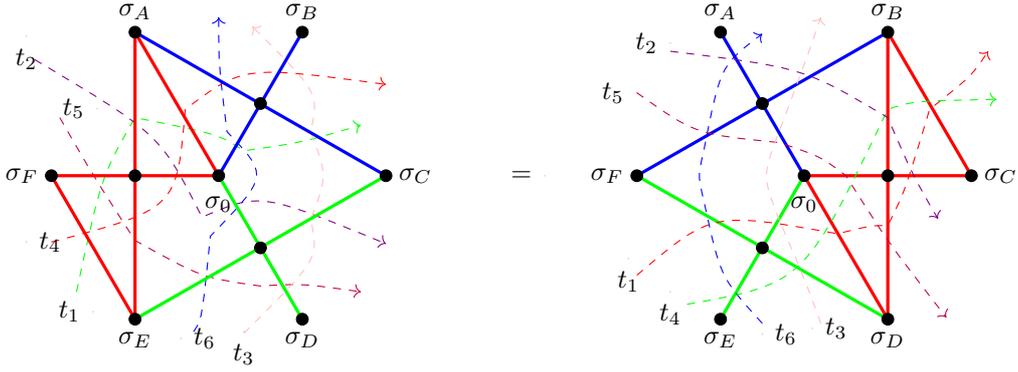

The star-star relation enables us to write IRF-type YBE by applying the star-star transformation in four steps from LHS to RHS as figured out in Fig.\ref{SSRYBE1}.

IRF-type Yang-Baxter equation of Fig.\ref{SSRYBE1} is indicated below and the explicit two Boltzmann weights on both sides are gauge factors that are annihilated and thus the star-star relation (\ref{gf}) and following identity (\ref{ssrybe}) assure us that the two models are identical.

\begin{align} 
\begin{aligned} 
\sum_{m_0\in \mathbb{Z}}\int d x_0 ~~
    \textcolor{red}{R_{t_{25}t_{41}}\left(\begin{array}{ccc}
        & \sigma_A & \\
\sigma_F         &  & \sigma_0 \\
&              \sigma_E     &
    \end{array}\right)}  
    \textcolor{blue}{R_{t_{63}t_{25}}\left(\begin{array}{cc}
    \sigma_0     & \sigma_C \\
\sigma_E         & \sigma_D
    \end{array}\right)}    \textcolor{green}{R_{t_{41}t_{63}}\left(\begin{array}{cc}
    \sigma_A     & \sigma_B \\
\sigma_0         & \sigma_C
    \end{array}\right)}  \makebox[4em]{}
      \\
\makebox[2em]{}
    =
    \sum_{m_0\in \mathbb{Z}} \int dx_0 ~~
        \textcolor{blue}{R_{t_{41}t_{63}}\left(\begin{array}{cc}
    \sigma_F     & \sigma_0 \\
\sigma_E         & \sigma_D
    \end{array}\right)}    \textcolor{green}{R_{t_{63}t_{25}}\left(\begin{array}{cc}
    \sigma_A     & \sigma_B \\
\sigma_F         & \sigma_0
    \end{array}\right)}    \textcolor{red}{R_{t_{25}t_{41}}\left(\begin{array}{ccc}
        & \sigma_B & \\
\sigma_0         &  & \sigma_C \\
&              \sigma_D     &
    \end{array}\right)}\:.
\end{aligned}
\end{align}\label{ssrybe}
 \makebox[4em]{}
      \\
\makebox[4em]{} 

The gauge/YBE correspondence gives a way to obtain many solutions to the IRF-type Yang-Baxter equation, see e.g. \cite{Yamazaki:2012cp,Gahramanov:2017idz,Gahramanov:2015cva,Yamazaki:2015voa,Kels:2017vbc}.

\section{Integrability property of Seiberg-like duality}
Here we discuss interesting identification from supersymmetric dualities by associating a quiver diagram to $\mathcal N=2$ supersymmetric gauge theories in three-dimension to integrability properties of lattice spin models in statistical mechanics. In this context, the integrability property refers to finding a solution to the Yang-Baxter equation. 
Quiver diagrams defining supersymmetric $\mathcal{N}=2$ quiver gauge theories and lattice of the spin models have the same structure. In both systems, there is a relationship named star-star relation on the spin model side and named Seiberg duality in the quiver gauge theory side. The gauge/YBE correspondence constructs a dictionary between these two fields of physics. 

By the dictionary of the correspondence, identifications of both sides will be explained, for more identifications\cite{Yamazaki:2013nra,Yamazaki:2018xbx}. 
We use vertices, edges, and interactions instead of loops, arrows, and bifundamental matter, respectively in the context of gauge/YBE correspondence, \cite{He:2004rn,Yamazaki:2008bt} for details on the quiver diagram.
In the schematic diagram, gauge groups such as $SU(2)$ or $U(1)\times U(1)$ will be represented on the vertices. Bifundamental matter sits on the edges and the arrow will play a role to specify fundamental (outgoing arrow) and antifundamental (incoming arrow) representations\footnote{In our study, there is no need to specify due to bifundamental matter content.} as rapidity lines to specify horizontal and vertical interaction types (explained in Sec.2 in details)\footnote{Additionally, we have not defined superpotential terms but in the existence of the term, it will be on the face of each square in the lattice.}. 

Supersymmetric quiver gauge theory shares the partition function with the corresponding integrable spin lattice model. Seiberg duality or star-star relation preserves the partition functions in the related sides. The Boltzmann weight of nearest-neighbor interaction is determined by contribution coming from chiral multiplets and vector multiplets corresponding to the Boltzmann weight of self-interaction.  

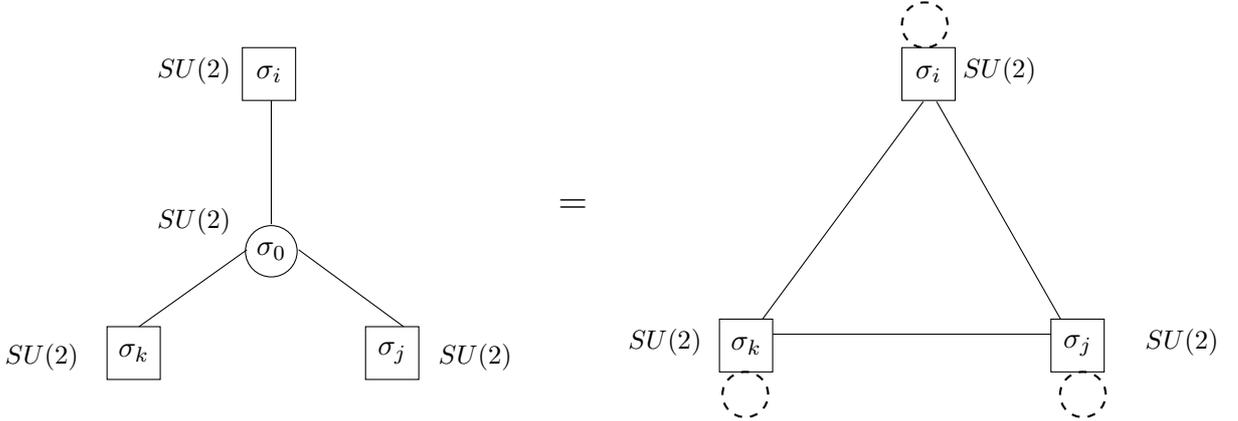
\begin{figure}[tbh]
\centering
\begin{tikzpicture}[scale=2]

\begin{scope}[xshift=10]
\draw (-2,1)--(-2,0.18);
\draw (-2.87,-0.5)--(-2.16,0.01);
\draw (-1.13,-0.5)--(-1.82,0.01);
\draw (-2,0) circle [radius=0.17] node {$\sigma_0$};
\draw (-1.84,1.35) rectangle ++(-0.35, -0.35) node[midway] {$\sigma_i$};
\draw (-2.73,-0.5) rectangle ++(-0.35, -0.35) node[midway] {$\sigma_k$};
\draw (-1.03,-0.5) rectangle ++(-0.35, -0.35) node[midway] {$\sigma_j$};

\fill[white!] (-2.2,1.2) circle (0.01pt)
node[left=0.05pt] {\color{black}{\small $SU(2)$}};
\fill[white!] (-3.2,-0.7) circle (0.01pt)
node[left=0.05pt] {\color{black}{\small $SU(2)$}};
\fill[white!] (-0.35,-0.7) circle (0.01pt)
node[left=0.05pt] {\color{black}{\small $SU(2)$}};
\fill[white!] (-2.2,0.2) circle (0.01pt)
node[left=0.05pt] {\color{black}{\small $SU(2)$}};

\end{scope}

\fill[white!] (0.5,0.3) circle (0.01pt)
node[left=0.05pt] {\Large\color{black}$=$};

\begin{scope}[xshift=-10]

\node[anchor=base] at (3.04,1.06) (i) {};
\node[anchor=base] at (1.86,-0.55) (k) {};
\node[anchor=base] at (3.95,-0.55) (j) {};

\path[-] (i) edge (k);
\path[-] (i) edge (j);
\path[-] (j) edge (k);

\filldraw[fill=white,draw=black] (3.2,1.35) rectangle ++(-0.35, -0.35) node[midway] {$\sigma_i$};
\filldraw[fill=white,draw=black] (2,-0.45) rectangle ++(-0.35, -0.35) node[midway] {$\sigma_k$};
\filldraw[fill=white,draw=black] (4.18,-0.45) rectangle ++(-0.35, -0.35) node[midway] {$\sigma_j$};

\draw[dashed,thick] (3,1.5) circle (0.15cm);
\draw[dashed,thick] (1.82,-0.95) circle (0.15cm);
\draw[dashed,thick] (4.04,-0.95) circle (0.15cm);

\fill[white!] (3.8,1.2) circle (0.01pt)
node[left=0.05pt] {\color{black}{\small $SU(2)$}};
\fill[white!] (5,-0.6) circle (0.01pt)
node[left=0.05pt] {\color{black}{\small $SU(2)$}};
\fill[white!] (1.6,-0.6) circle (0.01pt)
node[left=0.05pt] {\color{black}{\small $SU(2)$}};
\end{scope}

\end{tikzpicture}
\caption{Seiberg duality}
\label{str-seiberg}
\end{figure}
Note that, spectral lines (zig-zag paths) should be indicated in this figure and other figures performed in this section. They are the same as drawn on spin lattices, therefore, we skipped them. The reader can find the explicit expressions in \cite{Yamazaki:2018xbx}.

Then we represent the star-triangle relation (Fig.\ref{str-seiberg}) on the quiver diagram via the duality of partition functions. Simply, it is a change of variables as introduced in the following\footnote{ In this study, we are not interested in that aspect but one can organize the spectral parameters for discrete variables as shown in \cite{Spiridonov:2019uuw}.}
\begin{align}
    \begin{array}{c}
 x_1=+x_i-i\alpha_i\,,\quad x_3=+x_j-i\alpha_j\,,\quad x_5=+x_k-i\alpha_k\,,\\[0.3cm]
 x_2=-x_i-i\alpha_i\,,\quad x_4=-x_j-i\alpha_j\,,\quad x_6=-x_k-i\alpha_k\,,
\end{array}
\label{chngvs1}
\end{align}
and
\begin{align}
    m_1=m_i\,,\quad m_2=-m_i\,,\quad m_3=m_j\,,\quad m_4=-m_j\,,\quad m_5=m_k\,,\quad m_6=-m_k\,.
    \label{chngvs2}
\end{align}

In the supersymmetric gauge theory's side, flavor symmetry is broken from $SU(6)$ group to $SU(2) \times SU(2) \times SU(2)$ by adding a certain superpotential.
The centered circle represents the gauge symmetry group of the theory at the right-hand side while dashed circles represent the remaining mesons living in the adjoint representation of the $SU(2)$ flavor symmetries. The circles around flavor groups correspond to a spin-independent function in statistical mechanics which can be normalized as discussed in Sec.\ref{secstr}.

    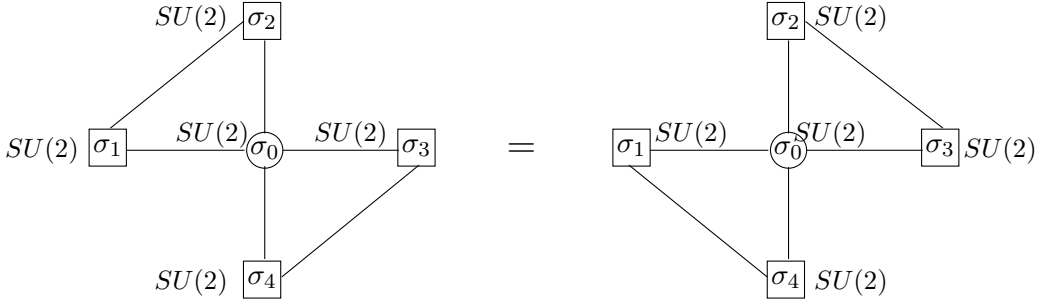
\begin{figure}[tbh]
\centering
\begin{tikzpicture}[scale=1.4]

\begin{scope}[xshift=10]
\draw (-4.3,0.25)--(-3.19,0.25);
\draw (-2.83,0.25)--(-1.74,0.25);
\draw (-3.2,1.47)--(-4.45,0.46);
\draw (-2.84,-0.95)--(-1.55,0.11);
\draw (-3,0.41)--(-3,1.29);
\draw (-3,0.10)--(-3,-0.78);
\draw (-3,0.25) circle [radius=0.17] node {$\sigma_0$};
\draw (-2.85,1.65) rectangle ++(-0.35, -0.35) node[midway] {$\sigma_2$};
\draw (-1.4,0.45) rectangle ++(-0.35, -0.35) node[midway] {$\sigma_3$};
\draw (-2.85,-0.8) rectangle ++(-0.35, -0.35) node[midway] {$\sigma_4$};
\draw (-4.3,0.45) rectangle ++(-0.35, -0.35) node[midway] {$\sigma_1$};
\end{scope}

\fill[white!] (-2.9,-1) circle (0.01pt)
node[left=0.05pt] {\color{black}{\small $SU(2)$}};
\fill[white!] (-2.9,1.5) circle (0.01pt)
node[left=0.05pt] {\color{black}{\small $SU(2)$}};
\fill[white!] (-4.3,0.25) circle (0.01pt)
node[left=0.05pt] {\color{black}{\small $SU(2)$}};
\fill[white!] (-2.7,0.4) circle (0.01pt)
node[left=0.05pt] {\color{black}{\small $SU(2)$}};
\fill[white!] (-1.4,0.4) circle (0.01pt)
node[left=0.05pt] {\color{black}{\small $SU(2)$}};

%\fill[white!] (-1.2,0.55) circle (0.01pt)
%node[left=0.05pt] {\color{black}$U(1)$};

\fill[white!] (0,0.25) circle (0.01pt)
node[left=0.05pt] {\Large\color{black}$=$};

\begin{scope}[xshift=150]
\draw (-4.3,0.25)--(-3.19,0.25);
\draw (-2.83,0.25)--(-1.74,0.25);
\draw (-2.84,1.47)--(-1.55,0.46);
\draw (-3.2,-0.95)--(-4.5,0.11);
\draw (-3,0.41)--(-3,1.29);
\draw (-3,0.09)--(-3,-0.78);
\draw (-3,0.25) circle [radius=0.17] node {$\sigma_0$};
\draw (-2.85,1.65) rectangle ++(-0.35, -0.35) node[midway] {$\sigma_2$};
\draw (-1.4,0.45) rectangle ++(-0.35, -0.35) node[midway] {$\sigma_3$};
\draw (-2.85,-0.8) rectangle ++(-0.35, -0.35) node[midway] {$\sigma_4$};
\draw (-4.3,0.45) rectangle ++(-0.35, -0.35) node[midway] {$\sigma_1$};
\end{scope}

\fill[white!] (3.3,-1) circle (0.01pt)
node[left=0.05pt] {\color{black}{\small $SU(2)$}};
\fill[white!] (3.3,1.5) circle (0.01pt)
node[left=0.05pt] {\color{black}{\small $SU(2)$}};
\fill[white!] (4.7,0.25) circle (0.01pt)
node[left=0.05pt] {\color{black}{\small $SU(2)$}};
\fill[white!] (3.1,0.4) circle (0.01pt)
node[left=0.05pt] {\color{black}{\small $SU(2)$}};
\fill[white!] (1.8,0.4) circle (0.01pt)
node[left=0.05pt] {\color{black}{\small $SU(2)$}};
%\fill[white!] (2.7,0.55) circle (0.01pt)
%node[left=0.05pt] {\color{black}$U(1)$};

\end{tikzpicture}
\caption{Diagrammatic picture of $\mathcal{N}=2$ Seiberg-like duality.}
\end{figure}\label{qief}

A similar change of variables (\ref{chngvs1}) and (\ref{chngvs2}) enables us to obtain a star-star relation depicted in Fig.\ref{qief}.
While constructing star-star relation, adding a superpotential breaks flavor symmetry from $SU(8)$ to $SU(2) \times SU(2) \times SU(2)\times SU(2)$ in the supersymmetry side.

Unlike the previous one, we have centered circles that are the gauge symmetry of the theories on both sides.

\begin{figure}[tbh]
\centering
\begin{tikzpicture}[scale=1.3]

\draw[-,very thick] (9,0)--(7,0);\draw[-,very thick] (8,1.73)--(8,-1.73);
\draw[-,very thick] (8,1.73)--(5,0);\draw[-,very thick] (6,1.73)--(7,0);
\draw[-,very thick] (8,-1.73)--(5,0);\draw[-,very thick] (6,-1.73)--(7,0);

\filldraw[fill=white,draw=black] (5.8,-1.93) rectangle (6.2,-1.53);

\draw[black] (6,-1.73) circle (0.01pt)
node[right=5 pt]{\footnotesize $SU(2)$}
;
\draw[black] (6,-1.73) circle (0.01pt)
node[]{\small $\sigma_E$}
;

\filldraw[fill=white,draw=black] (8.2,-1.93) rectangle (7.8,-1.53);

\draw[black] (8,-1.73) circle (0.01pt)
node[right=5 pt]{\footnotesize $SU(2)$}
;
\draw[black] (8,-1.73) circle (0.01pt)
node[]{\small $\sigma_D$}
;

\filldraw[fill=white,draw=black] (9.2,-0.2) rectangle (8.8,0.2);

\draw[black] (9,0) circle (0.01pt)
node[right=5 pt]{\footnotesize $SU(2)$}
;
\draw[black] (9,0) circle (0.01pt)
node[]{\small $\sigma_C$}
;

\filldraw[fill=white,draw=black] (5.8,1.93) rectangle (6.2,1.53);

\draw[black] (6,1.73) circle (0.01pt)
node[right=5 pt]{\footnotesize $SU(2)$}
;
\draw[black] (6,1.73) circle (0.01pt)
node[]{\small $\sigma_A$}
;

\filldraw[fill=white,draw=black] (8.2,1.93) rectangle (7.8,1.53);

\draw[black] (8,1.73) circle (0.01pt)
node[right=5 pt]{\footnotesize $SU(2)$}
;
\draw[black] (8,1.73) circle (0.01pt)
node[]{\small $\sigma_B$}
;

\filldraw[fill=white,draw=black] (4.8,-0.2) rectangle (5.2,0.2);

\draw[black] (5,0) circle (0.01pt)
node[right=5 pt]{\footnotesize $SU(2)$}
;
\draw[black] (5,0) circle (0.01pt)
node[]{\small $\sigma_F$}
;

\filldraw[fill=white,draw=black] (7,0) circle (6.0pt);
\draw[black] (7,0) circle (0.01pt)
node[above=5 pt]{\footnotesize $SU(2)$}
;
\draw[black] (7,0) circle (0.01pt)
node[]{\small $\sigma_0$};
\filldraw[fill=white,draw=black] (6.5,0.87) circle (6.0pt);
\draw[black] (6.5,0.87) circle (0.01pt)
node[right=5 pt]{\footnotesize $SU(2)$}
;
\draw[black] (6.5,0.87) circle (0.01pt)
node[]{\small $\sigma$};
\filldraw[fill=white,draw=black] (6.5,-0.87) circle (6.0pt);
\draw[black] (6.5,-0.87) circle (0.01pt)
node[right=5 pt]{\footnotesize $SU(2)$}
;
\draw[black] (6.5,-0.87) circle (0.01pt)
node[]{\small $\sigma$};
\filldraw[fill=white,draw=black] (8,0) circle (6.0pt);
\draw[black] (8,0) circle (0.01pt)
node[above=5 pt]{\footnotesize $SU(2)$}
;
\draw[black] (8,0) circle (0.01pt)
node[]{\small $\sigma$};

\draw[-,very thick] (-2,0)--(0,0);
\draw[-,very thick] (-1,1.73)--(-1,-1.73);
\draw[-,very thick] (-1,1.73)--(2,0);
\draw[-,very thick] (1,1.73)--(0,0);
\draw[-,very thick] (-1,-1.73)--(2,0);
\draw[-,very thick] (1,-1.73)--(0,0);

\filldraw[fill=white,draw=black] (-1.2,-1.93) rectangle (-0.8,-1.53);

\draw[black] (-1,-1.73) circle (0.01pt)
node[left=5 pt]{\footnotesize $SU(2)$}
;
\draw[black] (-1,-1.73) circle (0.01pt)
node[]{\small $\sigma_E$}
;

\filldraw[fill=white,draw=black] (1.2,-1.93) rectangle (0.8,-1.53);

\draw[black] (1,-1.73) circle (0.01pt)
node[left=5 pt]{\footnotesize $SU(2)$}
;
\draw[black] (1,-1.73) circle (0.01pt)
node[]{\small $\sigma_D$}
;

\filldraw[fill=white,draw=black] (2.2,-0.2) rectangle (1.8,0.2);

\draw[black] (2,0) circle (0.01pt)
node[left=5 pt]{\footnotesize $SU(2)$}
;
\draw[black] (2,0) circle (0.01pt)
node[]{\small $\sigma_C$}
;

\filldraw[fill=white,draw=black] (-1.2,1.93) rectangle (-0.8,1.53);

\draw[black] (-1,1.73) circle (0.01pt)
node[left=5 pt]{\footnotesize $SU(2)$}
;
\draw[black] (-1,1.73) circle (0.01pt)
node[]{\small $\sigma_A$}
;

\filldraw[fill=white,draw=black] (1.2,1.93) rectangle (0.8,1.53);

\draw[black] (1,1.73) circle (0.01pt)
node[left=5 pt]{\footnotesize $SU(2)$}
;
\draw[black] (1,1.73) circle (0.01pt)
node[]{\small $\sigma_B$}
;

\filldraw[fill=white,draw=black] (-2.2,-0.2) rectangle (-1.8,0.2);

\draw[black] (-2,0) circle (0.01pt)
node[left=5 pt]{\footnotesize $SU(2)$}
;
\draw[black] (-2,0) circle (0.01pt)
node[]{\small $\sigma_F$}
;

\filldraw[fill=white,draw=black] (0,0) circle (6.0pt);
\draw[black] (0,0) circle (0.01pt)
node[above=5 pt]{\footnotesize $SU(2)$}
;
\draw[black] (0,0) circle (0.01pt)
node[]{\small $\sigma_0$};

\filldraw[fill=white,draw=black] (0.5,0.87) circle (6.0pt);
\draw[black] (0.5,0.87) circle (0.01pt)
node[left=5 pt]{\footnotesize $SU(2)$}
;
\draw[black] (0.5,0.87) circle (0.01pt)
node[]{\small $\sigma$};

\filldraw[fill=white,draw=black] (0.5,-0.87) circle (6.0pt);
\draw[black] (0.5,-0.87) circle (0.01pt)
node[left=5 pt]{\footnotesize $SU(2)$}
;
\draw[black] (0.5,-0.87) circle (0.01pt)
node[]{\small $\sigma$};

\filldraw[fill=white,draw=black] (-1,0) circle (6.0pt);
\draw[black] (-1,0) circle (0.01pt)
node[above=5 pt]{\footnotesize $SU(2)$}
;
\draw[black] (-1,0) circle (0.01pt)
node[]{\small $\sigma$};

\draw[black] (3.7,0) circle (0.01pt)
node[left=1pt]{\color{black}\small $=$};

\end{tikzpicture}
\caption{Representation of the duality in quiver diagrams.}
\label{qirf}
\end{figure}
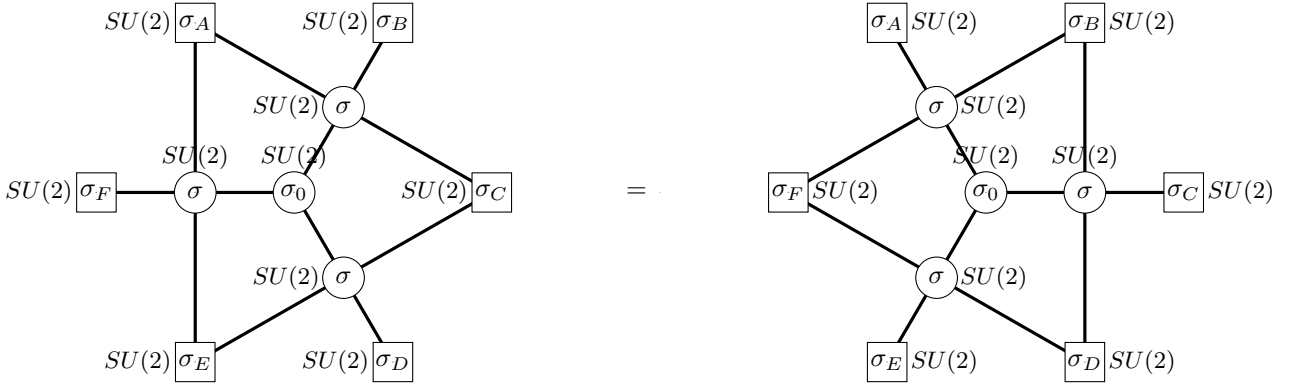

In the Fig.\ref{qirf}, we represent the IRF-type YBE performed in Fig.\ref{STRYBE} in  the quiver notation. 
 While the representation of circles corresponds to the gauge symmetry groups of a theory, the boxes correspond to flavor symmetry as figured in previous dualities\footnote{For more details and the corresponding representation of Fig.\ref{SSRYBE1} see \cite{Yamazaki:2018xbx}.}.

\section{Solutions to the star-star equation via partition functions on \texorpdfstring{$S_b^3/\mathbb{Z}_r$}{Sb3/Zr}}

\subsection{Solution by dual theories with \texorpdfstring{$SU(2)$}{SU(2)} gauge symmetry}\label{su2sol}

The hyperbolic hypergeometric solution to the star-triangle relation was constructed \cite{Gahramanov:2016ilb}.

	\begin{align} \nonumber
	\frac{1}{2r\sqrt{-\omega_1\omega_2}} \sum_{y=0}^{[ r/2 ]}\epsilon (y) \int _{-\infty}^{\infty} dz\frac{\prod_{i=1}^6
	\gamma^{(2)}(-i(a_i\pm z)-i\omega_1(u_i\pm y);-i\omega_1r,-i\omega)}
	{\gamma^{(2)}(\pm 2iz\pm i\omega_12y;-i\omega_1r,-i\omega)} \nonumber \\
	\times \frac{\gamma^{(2)}(-i(a_i\pm z)-i\omega_2(r-(u_i\pm y));-i\omega_2r,-i\omega)}
	{\gamma^{(2)}(\pm2iz-i\omega_2(r\pm2y);-i\omega_2r,-i\omega)} \nonumber \\ =\prod_{1\leq i<j\leq 6}\gamma^{(2)}(-i(a_i + a_j)-i\omega_1(u_i + u_j);-i\omega_1r,-i\omega) \nonumber \\ \times \gamma^{(2)}(-i(a_i + a_j)-i\omega_2(r-(u_i + u_j));-i\omega_2r,-i\omega)\; ,\label{sb3zr}
	\end{align}
where the balancing conditions are $\sum_{i=1}^6a_i=\omega_1+\omega_2$ and $\sum_{i=1}^6u_i=0$.  The function $\epsilon(y)$  takes $\epsilon(y)=2$ for all values except for $\epsilon(0)=\epsilon(\lfloor\frac{r}{2}\rfloor)=1$. 

One can introduce new variables $a_i  =-\alpha_i+x_{i} $ and $ a_{i+3}=-\alpha_i-x_{i} $ with the condition $u_i=-u_{i+3}$ for $i=1,2,3$  to define the following Boltzmann weight 
\begin{align}
    W_{\alpha}(x_i,x_j,u_i,u_j)=
    &\gamma^{(2)}(-i(-\alpha+x_i\pm x_j)-i\omega_1(u_i\pm u_j);-i\omega_{1}r,-i\omega)\nonumber \\&\times \gamma^{(2)}(-i(-\alpha+x_i\pm x_j)-i\omega_2(r-(u_i\pm u_j));-i\omega_2r,-i\omega)\nonumber \\&\times \gamma^{(2)}(-i(-\alpha-x_i\pm x_j)-i\omega_1(u_j\pm u_i);-i\omega_{1}r,-i\omega)\nonumber \\&\times \gamma^{(2)}(-i(-\alpha-x_i\pm x_j)-i\omega_2(r-(u_j\pm u_i));-i\omega_2r,-i\omega) \;.
\end{align}
Afterward, one can rewrite the integral identity (\ref{sb3zr}) in the form of the star-triangle equation.

The star-star relation for the same lattice spin model is the following integral identity

\begin{align} \nonumber 
      \sum_{y=0}^{[ r/2 ]}\epsilon (y) \int _{-\infty}^{\infty} \frac{\prod_{i=1}^8 \gamma^{(2)}(-i(a_i\pm z)-i\omega_1(u_i\pm y);-i\omega_1r,-i\omega)}{\gamma^{(2)}(\pm 2iz\pm i\omega_12y;-i\omega_1r,-i\omega)} \nonumber \\
     \times \frac{\gamma^{(2)}(-i(a_i\pm z)-i\omega_2(r-(u_i\pm y));-i\omega_2r,-i\omega)}{\gamma^{(2)}(\pm2iz-i\omega_2(r\pm2y);-i\omega_2r,-i\omega)} 
     \frac{dz}{r\sqrt{-\omega_1\omega_2}}
     \nonumber \\ 
     =
     \frac{\prod_{1\leq i<j\leq 4}\gamma^{(2)}(-i(a_i + a_j)-i\omega_1(u_i + u_j);-i\omega_1r,-i\omega) }
     {\prod_{5\leq i<j\leq 8}\gamma^{(2)}(-i(\Tilde{a}_i+\Tilde{a}_j)-i\omega_1(\Tilde{u}_i+\Tilde{u}_j);-i\omega_1r,-i\omega)  } \nonumber \\
    \times \frac{\prod_{1\leq i<j\leq 4}  \gamma^{(2)}(-i(a_i + a_j)-i\omega_2(r-(u_i + u_j));-i\omega_2r,-i\omega)}
     {\prod_{5\leq i<j\leq 8} \gamma^{(2)}(-i(\Tilde{a}_i+\Tilde{a}_j)-i\omega_2(r-(\Tilde{u}_i+\Tilde{u}_j));-i\omega_2r,-i\omega)} \nonumber \\
     \times   \sum_{m=0}^{[ r/2 ]}\epsilon (m) \int _{-\infty}^{\infty} \frac{\prod_{i=1}^8\gamma^{(2)}(-i(\Tilde{a}_i\pm x)-i\omega_1(\Tilde{u}_i\pm m);-i\omega_1r,-i\omega)}{\gamma^{(2)}(\pm 2ix\pm i\omega_12m;-i\omega_1r,-i\omega)} \nonumber \\
     \times \frac{\gamma^{(2)}(-i(\Tilde{a}_i\pm x)-i\omega_2(r-(\Tilde{u}_i\pm m));-i\omega_2r,-i\omega)}{\gamma^{(2)}(\pm2ix-i\omega_2(r\pm2m);-i\omega_2r,-i\omega)}
     \frac{dx}{r\sqrt{-\omega_1\omega_2}}
      \:,\label{newssr}
\end{align}

where the balancing conditions are $\sum_{i=1}^8a_i=2(\omega_1+\omega_2)$, $\sum_{i=1}^8u_i=0$,
and notations are identified as 
\begin{align}
\begin{aligned}
     \tilde{a}_i & =  a_i+s, & \tilde{u}_i & =u_i+p,  & \text{if} \;\;\; i=1,2,3,4 \:,
     \\ 
     \tilde{a}_i & =  a_i-s,  & \tilde{u}_i & = u_i-p, & \text{if} \;\;\; i=5,6,7,8,
\end{aligned}
\end{align}
where 
\begin{align}
\begin{aligned}
s & =\frac{1}{2}\left(\omega_1+\omega_2-\sum_{i=1}^4a_i\right)=\frac{1}{2}\left(-\omega_1-\omega_2+\sum_{i=5}^8a_i\right) \\
p & =-\frac{1}{2}\left(\sum_{i=1}^4u_i\right)=\frac{1}{2}\left(\sum_{i=5}^8u_i\right) \:.
\end{aligned}
\end{align}

In the case of $r=1$, the integral identity is appeared in \cite{Sarkissian:2020ipg}.

\subsection{Gauge symmetry breaking}

Here, we perform gauge symmetry breaking to complete the network represented in Fig.\ref{four}. Since we obtain the result \cite{Catak:2021coz} by our new star-star relation (\ref{newssr}), the gauge reduction will be seen as the way of the derivation of another hyperbolic hypergeometric solution to the star-star equation via gauge/YBE correspondence.

In our case the gauge symmetry breaking is a reduction from $SU(2)$ to $U(1)$ gauge symmetry group for two dual theories. In this procedure,  $SU(8)$ flavor symmetry group will be reduced to $SU(4)\times SU(4)$. 

We follow similar steps done \cite{Spiridonov:2010em,Bozkurt:2020gyy} as explained in Sec.\ref{gsb}

\begin{itemize}
    \item We change the boundaries from $\{-\infty,\infty \}$ to $\{0,\infty\}$ to make a change of variables $z\to z+\mu$ for left and $x\to x+\mu$ for right by using the symmetry property of the integrals (\ref{newssr}).
    \item We rewrite fugacities as $a_i$ with $a_i+\mu$ for $i=1,2,5,6$ and $a_i-\mu$ for  $i=3,4,7,8$ \:.
    \item As planned, change the variable $z$ to $z+\mu$, and $x$ to $x+\mu$ \:.
    \item After these manipulations, we take the limit $\mu\to \infty$, which physically corresponds to the fact that the mass of quarks is sent to infinity.
    \item Lastly, for the resulting identity, we relabel $a_{3,4}=b_{1,2}$, $a_{5,6}=a_{3,4}$ and $a_{7,8}=b_{3,4}$ together with $u_{3,4}=v_{1,2}$, $u_{5,6}=u_{3,4}$ and $u_{7,8}=v_{3,4}$ \:.
\end{itemize}

We should highlight here that all these steps preserve the balancing conditions on both sides and the resulting identity is

\begin{align}
  \sum_{y=0}^{[ r/2 ]}\epsilon (y) \int _{-\infty}^{\infty} \prod_{i=1}^4  \gamma^{(2)}(-i(a_i-z)-i\omega_1(u_i- y);-i\omega_1r,-i\omega) \nonumber \\
     \times  \gamma^{(2)}(-i(a_i-z)-i\omega_2(r-(u_i- y));-i\omega_2r,-i\omega) \nonumber \\
    \times  \gamma^{(2)}(-i(b_i+z)-i\omega_1(v_i+ y);-i\omega_1r,-i\omega) \nonumber
    \\
     \times  \gamma^{(2)}(-i(b_i+z)-i\omega_2(r-(v_i+y));-i\omega_2r,-i\omega)  
     \frac{dz}{r\sqrt{-\omega_1\omega_2}}\nonumber \\
     =\frac{e^{\frac{\pi i}{2}\sum_{i=1}^2(u_i-v_i)}}{e^{\frac{\pi i}{2}\sum_{i=3}^4(\tilde{u_i}-\tilde{v_i})}}
   \frac{\prod_{i,j=1}^2 \gamma^{(2)}(-i(a_i+ b_j)-i\omega_1(u_i+ v_j);-i\omega_1r,-i\omega) }{\prod_{i,j=3}^4 \gamma^{(2)}(-i(\tilde{a_i}+\tilde{b_j})-i\omega_1(\tilde{u_i}+\tilde{v_j});-i\omega_1r,-i\omega) } 
 \nonumber \\ \times  
     \frac{\prod_{i,j=1}^2  \gamma^{(2)}(-i(a_i+ b_j)-i\omega_2(r-(u_i+ v_j));-i\omega_2r,-i\omega)}{\prod_{i,j=3}^4  \gamma^{(2)}(-i(\tilde{a_i}+\tilde{b_j})-i\omega_2(r-(\tilde{u_i}+\tilde{v_j}));-i\omega_2r,-i\omega) } 
  \nonumber \\ \times  
     \sum_{m=0}^{[ r/2 ]}\epsilon (m) \int _{-\infty}^{\infty} \prod_{i=1}^4  \gamma^{(2)}(-i(\tilde{a_i}-x)-i\omega_1(\tilde{u_i}- m);-i\omega_1r,-i\omega) \nonumber \\
     \times  \gamma^{(2)}(-i(\tilde{a_i}-x)-i\omega_2(r-(\tilde{u_i}- m));-i\omega_2r,-i\omega) \nonumber \\
    \times  \gamma^{(2)}(-i(\tilde{b_i}+x)-i\omega_1(\tilde{v_i}+ m);-i\omega_1r,-i\omega) \nonumber
    \\
     \times  \gamma^{(2)}(-i(\tilde{b_i}+x)-i\omega_2(r-(\tilde{v_i}+m));-i\omega_2r,-i\omega) \frac{dx}{r\sqrt{-\omega_1\omega_2}} \label{u1ssr},
\end{align}
where the balancing conditions are $\sum_{i=1}^4a_i+b_i=2(\omega_1+\omega_2)$ and $\sum_{i=1}^4u_i+v_i=0$,
and we used the following notations
\begin{align}
\begin{aligned}
     \tilde{a}_i & =  a_i+s, & \tilde{b}_i & =  b_i+s, & \tilde{u}_i & =u_i+p, & \tilde{v}_i  & = v_i+p, & \text{if} \;\;\; i=1,2\:,
     \\ 
     \tilde{a}_i & =  a_i-s, & \tilde{b}_i & =  b_i-s, & \tilde{u}_i & = u_i-p, & \tilde{v}_i & =v_i-p, & \text{if} \;\;\; i=3,4 \:,
\end{aligned}
\end{align}
where 
\begin{align}
\begin{aligned}
s & =\frac{1}{2}(\omega_1+\omega_2-a_1-a_2-b_1-b_2)=\frac{1}{2}(-\omega_1-\omega_2+a_3+a_4+b_3+b_4)\:, \\
p & =-\frac{1}{2}(u_1+u_2+v_1+v_2)=\frac{1}{2}(u_3+u_4+v_3+v_4) \:.
\end{aligned}
\end{align}

The equation (\ref{u1ssr}) as a result and $r=1$ case\footnote{For more details, see \cite{Bult2007,Dimofte:2012pd}.} are consistent with \cite{Catak:2021coz,Sarkissian:2020ipg} and the gauge symmetry breaking for $r=1$ case can be found in \cite{Sarkissian:2020ipg}.

\subsection{Solution by dual theories with \texorpdfstring{$U(1)$}{U(1)} gauge symmetry}

By introducing new variables $a_i  =-\alpha_i+x_{i} $ and $ b_i=-\alpha_i-x_{i} $ with the condition $u_i=-v_i$ for $i=\overline{1,3}$,  one can rewrite the integral identity (\ref{u1str}) as the star-triangle equation \cite{Bozkurt:2020gyy, Sarkissian:2018ppc} with the following Boltzmann weight 
\begin{align}
    W_{\alpha}(x_i,x_j,u_i,u_j)=&e^{-\pi i(u_i+u_j)}\gamma^{(2)}(-i(-\alpha+x_i-x_j)-i\omega_1(u_i-u_j);-i\omega_{1}r,-i\omega)\nonumber \\&\times \gamma^{(2)}(-i(-\alpha+x_i-x_j)-i\omega_2(r-(u_i-u_j));-i\omega_2r,-i\omega)\nonumber \\&\times \gamma^{(2)}(-i(-\alpha-x_i+x_j)-i\omega_1(u_j-u_i);-i\omega_{1}r,-i\omega)\nonumber \\&\times \gamma^{(2)}(-i(-\alpha-x_i+x_j)-i\omega_2(r-(u_j-u_i));-i\omega_2r,-i\omega) \;.
\end{align}
The same Boltzmann weight holds for the (\ref{u1ssr}) by redefining the variables. As proposed, we obtain the star-star relation by a reduction of another star-star relation instead of using the star-triangle relation \cite{Catak:2021coz}.

\section{More results for beta integrals}

In this section, we mainly focus on mathematical identities which are obtained via reducing the number of flavors and they are consistent with the work \cite{Sarkissian:2020ipg} in the case of $r=1$.

\subsection{Reduction from \texorpdfstring{$N_f =6$}{Nf=6} to \texorpdfstring{$N_f =4$}{Nf=4}}

There was shown that the three-dimensional $\mathcal{N}=2$ theory with $N_f=6$ on $S^3_b$ can be reduced to $N_f=4$ theory \cite{Sarkissian:2020ipg}. We obtain similar equality for partition functions on $S^3_b/\mathbb{Z}_r$.\footnote{ One can see that our result will give the same integral identity in \cite{Sarkissian:2020ipg} by fixing $r=1$. At the same time, the limit $r\to \infty$, as in \cite{Eren:2019ibl}, will give similar results as obtained in \cite{Sarkissian:2020ipg} and \cite{Derkachov:2019ynh} so the question will help to clarify contradiction as discussed in \cite{Sarkissian:2020ipg}. } 

Now, we take the following limits in the duality (\ref{Sb3Zr})

\begin{align}
\begin{aligned}
    a_5 & \to \infty  \: , & a_6 & = \omega_1+\omega_2-\sum_{i=1}^5a_i \to -\infty  ~~~ \text{and} & u_5 & = u_6 \: .
\end{aligned}
\end{align}
We operate on the terms in the square parentheses
	\begin{align} \nonumber
	\frac{1}{2r\sqrt{-\omega_1\omega_2}} \sum_{y=0}^{[ r/2 ]}\epsilon (y) \int _{-\infty}^{\infty} dz\frac{\prod_{i=1}^4
	\gamma^{(2)}(-i(a_i\pm z)-i\omega_1(u_i\pm y);-i\omega_1r,-i\omega)}
	{\gamma^{(2)}(\pm 2iz\pm i\omega_12y;-i\omega_1r,-i\omega)} 
\nonumber \\\times
\frac{\gamma^{(2)}(-i(a_i\pm z)-i\omega_2(r-(u_i\pm y));-i\omega_2r,-i\omega)}
{\gamma^{(2)}(\pm2iz-i\omega_2(r\pm2y);-i\omega_2r,-i\omega)} 
     \nonumber \\	\times \Bigg[\prod_{i=5}^6
	\gamma^{(2)}(-i(a_i\pm z)-i\omega_1(u_i\pm y);-i\omega_1r,-i\omega)
\nonumber \\
	\times \gamma^{(2)}(-i(a_i\pm z)-i\omega_2(r-(u_i\pm y));-i\omega_2r,-i\omega)\Bigg]
	\nonumber \\ =\prod_{1\leq i<j\leq 4}\gamma^{(2)}(-i(a_i + a_j)-i\omega_1(u_i + u_j);-i\omega_1r,-i\omega) \nonumber \\ \times \gamma^{(2)}(-i(a_i + a_j)-i\omega_2(r-(u_i + u_j));-i\omega_2r,-i\omega)
	\nonumber \\ \times \Bigg[\prod_{i=1}^4\prod_{j=5}^6\gamma^{(2)}(-i(a_i + a_j)-i\omega_1(u_i + u_j);-i\omega_1r,-i\omega) \nonumber \\ \times \gamma^{(2)}(-i(a_i + a_j)-i\omega_2(r-(u_i + u_j));-i\omega_2r,-i\omega) 
\Bigg]	\nonumber \\ \times \gamma^{(2)}(-i(a_5 + a_6)-i\omega_1(u_5 + u_6);-i\omega_1r,-i\omega) 
	\nonumber \\ \times \gamma^{(2)}(-i(a_5 + a_6)-i\omega_2(r-(u_5 + u_6));-i\omega_2r,-i\omega)\; .
	\label{56red}
	\end{align}
	
By using the asymptotics (\ref{asymp}) of the hyperbolic gamma function	for the left-hand side 
	\begin{align}
	    & \lim_{a_5,-a_6\to \infty}\gamma^{(2)}(-i(a_5\pm z)-i\omega_1(u_5\pm y);-i\omega_1r,-i\omega) \nonumber \\
     &~~~~~~~~~~~~\times \gamma^{(2)}(-i(a_5\pm z)-i\omega_2(r-(u_5\pm y));-i\omega_2r,-i\omega)
     \nonumber \\
     &~~~~~~~~~~~~\times \gamma^{(2)}(-i(a_6\pm z)-i\omega_1(u_6\pm y);-i\omega_1r,-i\omega) \nonumber \\
     &~~~~~~~~~~~~\times \gamma^{(2)}(-i(a_6\pm z)-i\omega_2(r-(u_6\pm y));-i\omega_2r,-i\omega)
     \nonumber \\&=
     e^{\frac{\pi i}{2}\big(B_{2,2}(-i(a_5\pm z)-i\omega_1(u_5\pm y);-i\omega_1r,-i\omega)
     +B_{2,2}(-i(a_5\pm z)-i\omega_2(r-(u_5\pm y));-i\omega_2r,-i\omega)\big)}
     \nonumber \\ &\times 
     e^{-\frac{\pi i}{2}\big(+B_{2,2}(-i(a_6\pm z)-i\omega_1(u_6\pm y);-i\omega_1r,-i\omega)
   +B_{2,2}(-i(a_6\pm z)-i\omega_2(r-(u_6\pm y));-i\omega_2r,-i\omega)\big)}\:,
	\end{align}
	
and the terms at the right-hand side

\begin{align}
   & \lim_{a_5,-a_6\to \infty}
    \prod_{i=1}^4\gamma^{(2)}(-i(a_i + a_5)-i\omega_1(u_i + u_5);-i\omega_1r,-i\omega) 
    \nonumber \\  &~~~~~~~~~~~~\times
    \gamma^{(2)}(-i(a_i + a_5)-i\omega_2(r-(u_i + u_5));-i\omega_2r,-i\omega)
     \nonumber \\  &~~~~~~~~~~~~\times 
     \gamma^{(2)}(-i(a_i + a_6)-i\omega_1(u_i + u_6);-i\omega_1r,-i\omega) 
     \nonumber \\  &~~~~~~~~~~~~\times 
     \gamma^{(2)}(-i(a_i + a_6)-i\omega_2(r-(u_i + u_6));-i\omega_2r,-i\omega)
     \nonumber \\& =
     \prod_{i=1}^4e^{\frac{\pi i}{2}\big(B_{2,2}(-i(a_i + a_5)-i\omega_1(u_i + u_5);-i\omega_1r,-i\omega)
     +B_{2,2}(-i(a_i + a_5)-i\omega_2(r-(u_i + u_5));-i\omega_2r,-i\omega)} \big)
     \nonumber \\ &\times  e^{-\frac{\pi i}{2}\big(
     B_{2,2}(-i(a_i + a_6)-i\omega_1(u_i + u_6);-i\omega_1r,-i\omega)
     +B_{2,2}(-i(a_i + a_6)-i\omega_2(r-(u_i + u_6));-i\omega_2r,-i\omega)\big)}\:, 
  	\end{align}   
we reduce the global symmetries on both sides and obtain the duality described in Fig.\ref{nf4}.
	
\begin{figure}[H]
\centering
\begin{tikzpicture}[scale=1]

\filldraw[fill=black,draw=black] (-6.65,-1) 
%circle (1.2pt)
node[] {\color{black}  {\Large  $\mathcal{Z}^{\textbf{A}}_{\mathcal{S}^3_b/\mathbb{Z}_r}$}};

\filldraw[fill=black,draw=black] (-4,-0.25) 
%circle (1.2pt)
node[] {\color{black} $ Gauge=SU(2)$};

\filldraw[fill=black,draw=black] (-4,-0.75) 
%circle (1.2pt)
node[] {\color{black} $Four~ Chiral~ Multiplet$ };

\filldraw[fill=black,draw=black] (-4,-1.25) 
%circle (1.2pt)
node[] {\color{black} $ Global~ Symmetry:$};

\filldraw[fill=black,draw=black] (-4,-1.75) 
%circle (1.2pt)
node[] {\color{black} $ SU(4)\times U(1)$};

\draw[-,dashed,blue] (-6,0)--(-2,0);
\draw[-,dashed,blue] (-6,-2)--(-2,-2);
\draw[-,dashed,blue] (-6,0)--(-6,-2);
\draw[-,dashed,blue] (-2,-2)--(-2,0);

\filldraw[fill=black,draw=black] (-1.65,-1) 
%circle (1.2pt)
node[] {\color{black} \textbf{=}};

\filldraw[fill=black,draw=black] (-0.65,-1) 
%circle (1.2pt)
node[] {\color{black}  {\Large  $\mathcal{Z}^{\textbf{B}}_{\mathcal{S}^3_b/\mathbb{Z}_r}$}};

\filldraw[fill=black,draw=black] (2,-0.25) 
%circle (1.2pt)
node[] {\color{black} $No~Gauge$ };

\filldraw[fill=black,draw=black] (2,-0.75) 
%circle (1.2pt)
node[] {\color{black} $Six~ Free~ Mesons$ };

\filldraw[fill=black,draw=black] (2,-1.25) 
%circle (1.2pt)
node[] {\color{black} $ Global~ Symmetry:$};

\filldraw[fill=black,draw=black] (2,-1.75) 
%circle (1.2pt)
node[] {\color{black} $ SU(4)\times U(1)$};

\draw[-,dashed,green] (0,0)--(4,0);
\draw[-,dashed,green] (0,-2)--(4,-2);
\draw[-,dashed,green] (0,0)--(0,-2);
\draw[-,dashed,green] (4,-2)--(4,0);

\end{tikzpicture}
\caption{Ingredients of dual two theories. }
\label{nf4}
\end{figure}

As an integral identity, it is equivalent to

\begin{align} \nonumber
    \frac{1}{2r\sqrt{-\omega_1\omega_2}} \sum_{y=0}^{[ r/2 ]}\epsilon (y) \int _{-\infty}^{\infty} dz\frac{\prod_{i=1}^4\gamma^{(2)}(-i(a_i\pm z)-i\omega_1(u_i\pm y);-i\omega_1r,-i\omega)}{\gamma^{(2)}(\pm 2iz\pm i\omega_12y;-i\omega_1r,-i\omega)} \nonumber \\
     \times \frac{\gamma^{(2)}(-i(a_i\pm z)-i\omega_2(r-(u_i\pm y));-i\omega_2r,-i\omega)}{\gamma^{(2)}(\pm2iz-i\omega_2(r\pm2y);-i\omega_2r,-i\omega)} 
     \nonumber \\=\prod_{1\leq i<j\leq 4}\gamma^{(2)}(-i(a_i + a_j)-i\omega_1(u_i + u_j);-i\omega_1r,-i\omega) 
     \nonumber \\  \times \gamma^{(2)}(-i(a_i + a_j)-i\omega_2(r-(u_i + u_j));-i\omega_2r,-i\omega)
     \nonumber \\ \times 
     \gamma^{(2)}(-i(\omega_1+\omega_2-\sum_{i=1}^4a_i)-i\omega_1(u_i + u_j);-i\omega_1r,-i\omega) \nonumber \\ 
     \times \gamma^{(2)}(-i(\omega_1+\omega_2-\sum_{i=1}^4a_i )-i\omega_2(r-(u_i + u_j));-i\omega_2r,-i\omega)\; ,
\end{align}

where there is no balancing condition. One can have the same identity in \cite{Sarkissian:2020ipg} by fixing $r=1$ and the identity in the case\footnote{For more details, see \cite{Ruijsenaars2003,STOKMAN2005119}} of $r=1$ is discussed as a star-triangle relation in \cite{Kels:2018xge}.

\subsection{Reduction from \texorpdfstring{$N_f =4$}{Nf=4} to \texorpdfstring{$N_f =3$}{Nf=3}}   

The (\ref{newssr}) under the given limits and the fixed variables as done in \cite{Sarkissian:2020ipg} for $r=1$ case

\begin{align}
\begin{aligned}
    a_7 & \to \infty  \: , & a_8 & =2 ( \omega_1+\omega_2)-\sum_{i=1}^7a_i \to -\infty  ~~~ \text{and} & u_7 & = u_8\: ,
\end{aligned}
\end{align}
 turns to the following identity without the balancing condition 

\begin{align} \nonumber 
       \sum_{y=0}^{[ r/2 ]}\epsilon (y) \int _{-\infty}^{\infty} dz\frac{\prod_{i=1}^6 \gamma^{(2)}(-i(a_i\pm z)-i\omega_1(u_i\pm y);-i\omega_1r,-i\omega)}{\gamma^{(2)}(\pm 2iz\pm i\omega_12y;-i\omega_1r,-i\omega)} \nonumber \\
     \times \frac{\gamma^{(2)}(-i(a_i\pm z)-i\omega_2(r-(u_i\pm y));-i\omega_2r,-i\omega)}{\gamma^{(2)}(\pm2iz-i\omega_2(r\pm2y);-i\omega_2r,-i\omega)} 
     \nonumber \\ 
     =
     \frac{\gamma^{(2)}(-i(a_5+a_6)-i\omega_1(r-(-u_5-u_6));-i\omega_1r,-i\omega)}
     {\gamma^{(2)}(-i(\sum_{i=1}^6a_i-\omega_1-\omega_2)-i\omega_1(\sum_{i=1}^6u_i);-i\omega_1r,-i\omega)}
     \nonumber \\
\times     \frac{\gamma^{(2)}(-i(a_5+a_6)-i\omega_2(-u_5-u_6);-i\omega_2r,-i\omega)}
     {\gamma^{(2)}(-i(\sum_{i=1}^6a_i-\omega_1-\omega_2)-i\omega_2(r-\sum_{i=1}^6u_i);-i\omega_2r,-i\omega)} \nonumber \\
\times     \prod_{1\leq i<j\leq 4}\gamma^{(2)}(-i(a_i + a_j)-i\omega_1(u_i + u_j);-i\omega_1r,-i\omega) 
     \nonumber \\
     \times \gamma^{(2)}(-i(a_i + a_j)-i\omega_2(r-(u_i + u_j));-i\omega_2r,-i\omega)
     \nonumber \\ 
     \times  \sum_{m=0}^{[ r/2 ]}\epsilon (m) \int _{-\infty}^{\infty} dx\frac{\prod_{i=1}^6\gamma^{(2)}(-i(\Tilde{a}_i\pm x)-i\omega_1(\Tilde{u}_i\pm m);-i\omega_1r,-i\omega)}{\gamma^{(2)}(\pm 2ix\pm i\omega_12m;-i\omega_1r,-i\omega)} \nonumber \\
     \times \frac{\gamma^{(2)}(-i(\Tilde{a}_i\pm x)-i\omega_2(r-(\Tilde{u}_i\pm m));-i\omega_2r,-i\omega)}{\gamma^{(2)}(\pm2ix-i\omega_2(r\pm2m);-i\omega_2r,-i\omega)}
       \:,
\end{align}

where 
\begin{align}
\begin{aligned}
     \tilde{a}_i & =  a_i+s, & \tilde{u}_i & =u_i+p, &  \text{if} \;\;\; i=1,2,3,4\:,
     \\ 
     \tilde{a}_i & =  a_i-s, & \tilde{u}_i & = u_i-p, & \text{if} \;\;\; i=5,6\:,
\end{aligned}
\end{align}
with 
\begin{align}
\begin{aligned}
s & =\frac{1}{2}(\omega_1+\omega_2-a_1-a_2-a_3-a_4)\:, \\
p & =-\frac{1}{2}(u_1+u_2+u_3+u_4) \:.
\end{aligned}
\end{align}

\subsection{Reduction from \texorpdfstring{$N_f =4$}{Nf=4} to \texorpdfstring{$N_f =3$}{Nf=3}}

After applying the following operations
\begin{align}
\begin{aligned}
    a_4 & \to \infty  \: , & b_4 & = \omega_1+\omega_2-\sum_{i=1}^3(a_i+b_i) \to -\infty  ~~~ \text{and} & u_4 & = v_4\:, 
\end{aligned}
\end{align}

the integral identity (\ref{u1ssr}) reduces to the equality

\begin{align}
  \sum_{y=0}^{[ r/2 ]}\epsilon (y) \int _{-\infty}^{\infty} e^{\pi i A} \prod_{i=1}^3  \gamma^{(2)}(-i(a_i-x)-i\omega_1(u_i- y);-i\omega_1r,-i\omega) \nonumber \\
     \times  \gamma^{(2)}(-i(a_i-x)-i\omega_2(r-(u_i- y));-i\omega_2r,-i\omega) \nonumber \\
    \times  \gamma^{(2)}(-i(b_i+x)-i\omega_1(v_i+ y);-i\omega_1r,-i\omega) \nonumber
    \\
     \times  \gamma^{(2)}(-i(b_i+x)-i\omega_2(r-(v_i+y));-i\omega_2r,-i\omega)  
     \frac{dx}{r\sqrt{-\omega_1\omega_2}}\nonumber \\
     = e^{\frac{\pi i C}{2}}
   \frac{\prod_{i,j=1}^2 \gamma^{(2)}(-i(a_i+ b_j)-i\omega_1(u_i+ v_j);-i\omega_1r,-i\omega) }{ \gamma^{(2)}(-i(\tilde{a_3}+\tilde{b_3})-i\omega_1(\tilde{u_3}+\tilde{v_3});-i\omega_1r,-i\omega) } 
 \nonumber \\ \times  
     \frac{\prod_{i,j=1}^2  \gamma^{(2)}(-i(a_i+ b_j)-i\omega_2(r-(u_i+ v_j));-i\omega_2r,-i\omega)}{   \gamma^{(2)}(-i(\tilde{a_3}+\tilde{b_3})-i\omega_2(r-(\tilde{u_3}+\tilde{v_3}));-i\omega_2r,-i\omega) } 
  \nonumber \\ \times  
 \gamma^{(2)}(-i(-\omega+\sum_{i=1}^3a_i+b_i)-i\omega_1(r+\sum_{i=1}^3\tilde{u_i}+\tilde{v_i});-i\omega_1r,-i\omega) 
 \nonumber \\ \times  
 \gamma^{(2)}(-i(-\omega+\sum_{i=1}^3a_i+b_i)-i\omega_2(-\sum_{i=1}^3\tilde{u_i}+\tilde{v_i}));-i\omega_2r,-i\omega) 
  \nonumber \\ \times
     \sum_{m=0}^{[ r/2 ]}\epsilon (m) \int _{-\infty}^{\infty} e^{\pi i B} \prod_{i=1}^3  \gamma^{(2)}(-i(\tilde{a_i}-z)-i\omega_1(\tilde{u_i}- m);-i\omega_1r,-i\omega) \nonumber \\
     \times  \gamma^{(2)}(-i(\tilde{a_i}-z)-i\omega_2(r-(\tilde{u_i}- m));-i\omega_2r,-i\omega) \nonumber \\
    \times  \gamma^{(2)}(-i(\tilde{b_i}+z)-i\omega_1(\tilde{v_i}+ m);-i\omega_1r,-i\omega) \nonumber
    \\
     \times  \gamma^{(2)}(-i(\tilde{b_i}+z)-i\omega_2(r-(\tilde{v_i}+m));-i\omega_2r,-i\omega) \frac{dz}{r\sqrt{-\omega_1\omega_2}} \:,
     \label{son}
\end{align}
where we have the following exponents
\begin{align}
\begin{aligned}
     &A =  x\frac{(\omega-\sum_{i=1}^3a_i+b_i )}{(-i\omega_1)(-i\omega_2)r} +\frac{m}{r} (r+\sum_{i=1}^3u_i+v_i),
     \\& 
     B =  z\frac{(-\omega-\tilde{a}_3-\tilde{b}_3 +\sum_{i=1}^2a_i+b_i )}{(-i\omega_1)(-i\omega_2)r} +\frac{y}{r} (-r+\tilde{u}_3+\tilde{v}_3 -\sum_{i=1}^2(u_i+v_i)),
     \\& 
     C=  \frac{(a_3-b_3)
    (\omega-\sum_{i=1}^2a_i+b_i)
}{(-i\omega_1)(-i\omega_2)r}+ \sum_{i=1}^2\frac{u_i+v_i}{r}+\sum_{i=1}^2(u_i-v_i)-2(\tilde{u}_3-\tilde{v}_3) \:,
\end{aligned}
\end{align}

and the notations are 
\begin{align}
\begin{aligned}
     \tilde{a}_i & =  a_i+s, & \tilde{b}_i & =  b_i+s, & \tilde{u}_i & =u_i+p, & \tilde{v}_i  & = v_i+p, & \text{if} \;\;\; i=1,2 \:,
     \\ 
     \tilde{a}_i & =  a_i-s, & \tilde{b}_i & =  b_i-s, & \tilde{u}_i & = u_i-p, & \tilde{v}_i & =v_i-p, & \text{if} \;\;\; i=3 \:,
\end{aligned}
\end{align}
with
\begin{align}
\begin{aligned}
s & =\frac{1}{2}(\omega_1+\omega_2-a_1-a_2-b_1-b_2) \:, \\
p & =-\frac{1}{2}(u_1+u_2+v_1+v_2) \:.
\end{aligned}
\end{align}

The result (\ref{son}) is also consistent for $r=1$ case appeared in \cite{Sarkissian:2020ipg}.

\section{Conclusions}

In this paper, a new hyperbolic solution to the star-star equation is constructed via the gauge/YBE correspondence. The gauge symmetry breaking method gave the equality of dual theories and the result is already written as the star-star relation in \cite{Catak:2021coz}. Hence, various connections between integrable lattice spin models and supersymmetric gauge theories are enhanced and proved in the correspondence.

On the other hand, the new solution can be also derived by the hyperbolic limit \cite{Gahramanov:2016ilb} of the Lens elliptic solution \cite{Kels:2015bda}. The limit will correspond to the dimensional reduction of the four-dimensional $\mathcal{N}=1$ supersymmetric gauge theories to the three-dimensional $\mathcal{N}=2$ theories.

Another way of constructing the star-star relation is the use of Bailey pair construction see \cite{Gahramanov:2015cva} and \cite{Gahramanov:2021pgu} of the star-triangle relation.

It would be interesting to obtain an Euler gamma solution to the star-star equation and other beta integrals in \cite{Sarkissian:2020ipg} by limiting $r\to \infty$ in our results. The physical interpretation of the resulting integral identities will be dual theories on $S^2$ as in \cite{Eren:2019ibl}.

%%%%%%%%%%%%%%%%%%%%%%%%%%%%%%%%%%%%%%%%%%%%%%%%%%%%%%%%%%%%%%%
\section*{Acknowledgments}
%%%%%%%%%%%%%%%%%%%%%%%%%%%%%%%%%%%%%%%%%%%%%%%%%%%%%%%%%%%%%%%
We thank Ilmar Gahramanov for suggesting the problem, reading the manuscript, and
his many insightful teachings on supersymmetric partition functions and exactly solvable
models. 
We are also grateful to Erdal Catak, Ege Aktener, and Taha Ayfer for their valuable discussions.  Mustafa Mullahasanoglu is supported by the 1002-TUBITAK Quick Support Program under grant number 121F413. 
Nuri Tas is supported by the 2209-TUBITAK National/International Research Projects Fellowship Programme for Undergraduate Students under grant number 1919B012005092.

\appendix

\bibliographystyle{utphys}
\bibliography{refYBE}

\providecommand{\href}[2]{#2}\begingroup\raggedright\begin{thebibliography}{10}

\bibitem{Baxter:1982zz}
R.~J. Baxter, {\em Exactly {S}olved {M}odels in {S}tatistical {M}echanics}.
\newblock Academic, London,
1982.
\newblock
%%CITATION = INSPIRE-1120339;%%.

\bibitem{baxter:1997ssr}
R.~J. Baxter, ``Star-triangle and star-star relations in statistical
  mechanics,'' \href{http://dx.doi.org/10.1142/S0217979297000058}{{\em
  International Journal of Modern Physics B} {\bf 11} (1997) no.~01n02,
  27--37}.

\bibitem{Bazhanov:2013bh}
V.~V. Bazhanov, A.~P. Kels, and S.~M. Sergeev, ``{Comment on star-star
  relations in statistical mechanics and elliptic gamma-function identities},''
  \href{http://dx.doi.org/10.1088/1751-8113/46/15/152001}{{\em J. Phys.} {\bf
  A46} (2013)  152001},
\href{http://arxiv.org/abs/1301.5775}{{\tt arXiv:1301.5775 [math-ph]}}.
%%CITATION = ARXIV:1301.5775;%%.

\bibitem{Gahramanov:2016ilb}
I.~Gahramanov and A.~P. Kels, ``{The star-triangle relation, lens partition
  function, and hypergeometric sum/integrals},''
  \href{http://dx.doi.org/10.1007/JHEP02(2017)040}{{\em JHEP} {\bf 02} (2017)
  040},
\href{http://arxiv.org/abs/1610.09229}{{\tt arXiv:1610.09229 [math-ph]}}.
%%CITATION = ARXIV:1610.09229;%%.

\bibitem{Bozkurt:2020gyy}
D.~N. Bozkurt, I.~Gahramanov, and M.~Mullahasanoglu, ``{Lens partition
  function, pentagon identity, and star-triangle relation},''
  \href{http://dx.doi.org/10.1103/PhysRevD.103.126013}{{\em Phys. Rev. D} {\bf
  103} (2021) no.~12, 126013}, \href{http://arxiv.org/abs/2009.14198}{{\tt
  arXiv:2009.14198 [hep-th]}}.

\bibitem{Catak:2021coz}
E.~Catak, I.~Gahramanov, and M.~Mullahasanoglu, ``{Hyperbolic and trigonometric
  hypergeometric solutions to the star-star equation},''
  \href{http://arxiv.org/abs/2107.06880}{{\tt arXiv:2107.06880 [hep-th]}}.

\bibitem{Gahramanov:2017ysd}
I.~Gahramanov and S.~Jafarzade, ``{Integrable lattice spin models from
  supersymmetric dualities},''
  \href{http://dx.doi.org/10.1134/S1547477118060079}{{\em Phys. Part. Nucl.
  Lett.} {\bf 15} (2018) no.~6, 650--667},
\href{http://arxiv.org/abs/1712.09651}{{\tt arXiv:1712.09651 [math-ph]}}.
%%CITATION = ARXIV:1712.09651;%%.

\bibitem{Yamazaki:2018xbx}
M.~Yamazaki, ``{Integrability as Duality: the Gauge/YBE Correspondence},''
\href{http://arxiv.org/abs/1808.04374}{{\tt arXiv:1808.04374 [hep-th]}}.
%%CITATION = ARXIV:1808.04374;%%.

\bibitem{Spiridonov:2010em}
V.~P. Spiridonov, ``{Elliptic beta integrals and solvable models of statistical
  mechanics},'' {\em Contemp. Math.} {\bf 563} (2012)  181--211,
\href{http://arxiv.org/abs/1011.3798}{{\tt arXiv:1011.3798 [hep-th]}}.
%%CITATION = ARXIV:1011.3798;%%.

\bibitem{Kels:2015bda}
A.~P. Kels, ``{New solutions of the star–triangle relation with discrete and
  continuous spin variables},''
  \href{http://dx.doi.org/10.1088/1751-8113/48/43/435201}{{\em J. Phys.} {\bf
  A48} (2015) no.~43, 435201},
\href{http://arxiv.org/abs/1504.07074}{{\tt arXiv:1504.07074 [math-ph]}}.
%%CITATION = ARXIV:1504.07074;%%.

\bibitem{Gahramanov:2015cva}
I.~Gahramanov and V.~P. Spiridonov, ``{The star-triangle relation and 3d
  superconformal indices},''
  \href{http://dx.doi.org/10.1007/JHEP08(2015)040}{{\em JHEP} {\bf 08} (2015)
  040},
\href{http://arxiv.org/abs/1505.00765}{{\tt arXiv:1505.00765 [hep-th]}}.
%%CITATION = ARXIV:1505.00765;%%.

\bibitem{Yamazaki:2013nra}
M.~Yamazaki, ``{New Integrable Models from the Gauge/YBE Correspondence},''
  \href{http://dx.doi.org/10.1007/s10955-013-0884-8}{{\em J. Statist. Phys.}
  {\bf 154} (2014)  895},
\href{http://arxiv.org/abs/1307.1128}{{\tt arXiv:1307.1128 [hep-th]}}.
%%CITATION = ARXIV:1307.1128;%%.

\bibitem{Kels:2017vbc}
A.~P. Kels and M.~Yamazaki, ``{Lens elliptic gamma function solution of the
  Yang–Baxter equation at roots of unity},''
  \href{http://dx.doi.org/10.1088/1742-5468/aaa8fd}{{\em J. Stat. Mech.} {\bf
  1802} (2018) no.~2, 023108},
\href{http://arxiv.org/abs/1709.07148}{{\tt arXiv:1709.07148 [math-ph]}}.
%%CITATION = ARXIV:1709.07148;%%.

\bibitem{Gahramanov:2015tta}
I.~Gahramanov, ``{Mathematical structures behind supersymmetric dualities},''
  \href{http://dx.doi.org/10.5817/AM2015-5-273}{{\em Archivum Math.} {\bf 51}
  (2015)  273--286},
\href{http://arxiv.org/abs/1505.05656}{{\tt arXiv:1505.05656 [math-ph]}}.
%%CITATION = ARXIV:1505.05656;%%.

\bibitem{Gahramanov:2022qge}
I.~Gahramanov, ``{Integrability from supersymmetric duality: a short review},''
  \href{http://arxiv.org/abs/2201.00351}{{\tt arXiv:2201.00351 [hep-th]}}.

\bibitem{Spiridonov:2019kto}
V.~P. Spiridonov, ``{Superconformal Indices, Seiberg Dualities and Special
  Functions},'' \href{http://dx.doi.org/10.1134/S1063779620040681}{{\em Phys.
  Part. Nucl.} {\bf 51} (2020) no.~4, 508--513},
  \href{http://arxiv.org/abs/1912.11514}{{\tt arXiv:1912.11514 [hep-th]}}.

\bibitem{Tachikawa:2017byo}
Y.~Tachikawa, ``{On 'categories' of quantum field theories},'' in {\em
  {International Congress of Mathematicians}}.
\newblock 12, 2017.
\newblock \href{http://arxiv.org/abs/1712.09456}{{\tt arXiv:1712.09456
  [math-ph]}}.

\bibitem{Spiridonov:2009za}
V.~P. Spiridonov and G.~S. Vartanov, ``{Elliptic Hypergeometry of
  Supersymmetric Dualities},''
  \href{http://dx.doi.org/10.1007/s00220-011-1218-9}{{\em Commun. Math. Phys.}
  {\bf 304} (2011)  797--874}, \href{http://arxiv.org/abs/0910.5944}{{\tt
  arXiv:0910.5944 [hep-th]}}.

\bibitem{Spiridonov2014}
V.~P. Spiridonov and G.~S. Vartanov, ``Elliptic hypergeometry of supersymmetric
  dualities ii. orthogonal groups, knots, and vortices,''
  \href{http://dx.doi.org/10.1007/s00220-013-1861-4}{{\em Communications in
  Mathematical Physics} {\bf 325} (2014) no.~2, 421--486},
\href{http://arxiv.org/abs/1107.5788}{{\tt arXiv:1107.5788 [hep-th]}}.
%%arXiv:1107.5788%%.

\bibitem{Krattenthaler:2011da}
C.~Krattenthaler, V.~P. Spiridonov, and G.~S. Vartanov, ``{Superconformal
  indices of three-dimensional theories related by mirror symmetry},''
  \href{http://dx.doi.org/10.1007/JHEP06(2011)008}{{\em JHEP} {\bf 06} (2011)
  008},
\href{http://arxiv.org/abs/1103.4075}{{\tt arXiv:1103.4075 [hep-th]}}.
%%CITATION = ARXIV:1103.4075;%%.

\bibitem{Gahramanov:gka}
I.~B. Gahramanov and G.~S. Vartanov,
  \href{http://dx.doi.org/10.1142/9789814449243_0076}{``{Superconformal indices
  and partition functions for supersymmetric field theories},''} in {\em
  {XVIIth Intern. Cong. Math. Phys. 695-703 (2013)}}.
\newblock 2013.
\newblock
\href{http://arxiv.org/abs/1310.8507}{{\tt arXiv:1310.8507 [hep-th]}}.
\newblock
%%CITATION = ARXIV:1310.8507;%%.

\bibitem{Dolan:2011rp}
F.~A.~H. Dolan, V.~P. Spiridonov, and G.~S. Vartanov, ``{From 4d superconformal
  indices to 3d partition functions},''
  \href{http://dx.doi.org/10.1016/j.physletb.2011.09.007}{{\em Phys. Lett.}
  {\bf B704} (2011)  234--241},
\href{http://arxiv.org/abs/1104.1787}{{\tt arXiv:1104.1787 [hep-th]}}.
%%CITATION = ARXIV:1104.1787;%%.

\bibitem{Bazhanov:2011mz}
V.~V. Bazhanov and S.~M. Sergeev, ``{Elliptic gamma-function and multi-spin
  solutions of the Yang-Baxter equation},''
  \href{http://dx.doi.org/10.1016/j.nuclphysb.2011.10.032}{{\em Nucl. Phys.}
  {\bf B856} (2012)  475--496},
\href{http://arxiv.org/abs/1106.5874}{{\tt arXiv:1106.5874 [math-ph]}}.
%%CITATION = ARXIV:1106.5874;%%.

\bibitem{Kels:2017toi}
A.~P. Kels and M.~Yamazaki, ``{Elliptic hypergeometric sum/integral
  transformations and supersymmetric lens index},''
  \href{http://dx.doi.org/10.3842/SIGMA.2018.013}{{\em SIGMA} {\bf 14} (2018)
  013}, \href{http://arxiv.org/abs/1704.03159}{{\tt arXiv:1704.03159
  [math-ph]}}.

\bibitem{Rains:2003}
E.~M. Rains, ``Transformations of elliptic hypergometric integrals,''.
  \url{https://arxiv.org/abs/math/0309252}.

\bibitem{Sarkissian:2020ipg}
G.~A. Sarkissian and V.~P. Spiridonov, ``{The Endless Beta Integrals},''
  \href{http://dx.doi.org/10.3842/SIGMA.2020.074}{{\em SIGMA} {\bf 16} (2020)
  074}, \href{http://arxiv.org/abs/2005.01059}{{\tt arXiv:2005.01059
  [math-ph]}}.

\bibitem{Khmelnitsky:2009vc}
A.~Khmelnitsky, ``{Interpreting multiple dualities conjectured from
  superconformal index identities},''
  \href{http://dx.doi.org/10.1007/JHEP03(2010)065}{{\em JHEP} {\bf 03} (2010)
  065},
\href{http://arxiv.org/abs/0912.4523}{{\tt arXiv:0912.4523 [hep-th]}}.
%%CITATION = ARXIV:0912.4523;%%.

\bibitem{van2007hyperbolic}
F.~van~de Bult {\em et al.}, ``Hyperbolic hypergeometric functions,'' {\em
  Ph.D. Thesis, University of Amsterdam, Amsterdam Netherlands} (2007)  .

\bibitem{Andersen:2014aoa}
J.~E. Andersen and R.~Kashaev, ``{Complex Quantum Chern-Simons},''
  \href{http://arxiv.org/abs/1409.1208}{{\tt arXiv:1409.1208 [math.QA]}}.

\bibitem{Faddeev:1995nb}
L.~Faddeev, ``{Discrete Heisenberg-Weyl group and modular group},''
  \href{http://dx.doi.org/10.1007/BF01872779}{{\em Lett. Math. Phys.} {\bf 34}
  (1995)  249--254}, \href{http://arxiv.org/abs/hep-th/9504111}{{\tt
  arXiv:hep-th/9504111}}.

\bibitem{woronowicz2000quantum}
S.~Woronowicz, ``Quantum exponential function,''
  \href{http://dx.doi.org/10.1142/S0129055X00000344}{{\em Reviews in
  Mathematical Physics} {\bf 12} (2000) no.~06, 873--920}.

\bibitem{Nieri:2015yia}
F.~Nieri and S.~Pasquetti, ``{Factorisation and holomorphic blocks in 4d},''
  \href{http://dx.doi.org/10.1007/JHEP11(2015)155}{{\em JHEP} {\bf 11} (2015)
  155},
\href{http://arxiv.org/abs/1507.00261}{{\tt arXiv:1507.00261 [hep-th]}}.
%%CITATION = ARXIV:1507.00261;%%.

\bibitem{Imamura:2011wg}
Y.~Imamura and D.~Yokoyama, ``{N=2 supersymmetric theories on squashed
  three-sphere},'' \href{http://dx.doi.org/10.1103/PhysRevD.85.025015}{{\em
  Phys. Rev.} {\bf D85} (2012)  025015},
\href{http://arxiv.org/abs/1109.4734}{{\tt arXiv:1109.4734 [hep-th]}}.
%%CITATION = ARXIV:1109.4734;%%.

\bibitem{Closset:2012ru}
C.~Closset, T.~T. Dumitrescu, G.~Festuccia, and Z.~Komargodski,
  ``{Supersymmetric Field Theories on Three-Manifolds},''
  \href{http://dx.doi.org/10.1007/JHEP05(2013)017}{{\em JHEP} {\bf 05} (2013)
  017},
\href{http://arxiv.org/abs/1212.3388}{{\tt arXiv:1212.3388 [hep-th]}}.
%%CITATION = ARXIV:1212.3388;%%.

\bibitem{Seiberg:1994pq}
N.~Seiberg, ``{Electric - magnetic duality in supersymmetric nonAbelian gauge
  theories},'' \href{http://dx.doi.org/10.1016/0550-3213(94)00023-8}{{\em Nucl.
  Phys.} {\bf B435} (1995)  129--146},
\href{http://arxiv.org/abs/hep-th/9411149}{{\tt arXiv:hep-th/9411149
  [hep-th]}}.
%%CITATION = HEP-TH/9411149;%%.

\bibitem{Intriligator:1996ex}
K.~A. Intriligator and N.~Seiberg, ``{Mirror symmetry in three-dimensional
  gauge theories},'' \href{http://dx.doi.org/10.1016/0370-2693(96)01088-X}{{\em
  Phys. Lett. B} {\bf 387} (1996)  513--519},
  \href{http://arxiv.org/abs/hep-th/9607207}{{\tt arXiv:hep-th/9607207}}.

\bibitem{Aharony:1997bx}
O.~Aharony, A.~Hanany, K.~A. Intriligator, N.~Seiberg, and M.~J. Strassler,
  ``{Aspects of N=2 supersymmetric gauge theories in three-dimensions},''
  \href{http://dx.doi.org/10.1016/S0550-3213(97)00323-4}{{\em Nucl. Phys. B}
  {\bf 499} (1997)  67--99}, \href{http://arxiv.org/abs/hep-th/9703110}{{\tt
  arXiv:hep-th/9703110}}.

\bibitem{Kapustin:2010xq}
A.~Kapustin, B.~Willett, and I.~Yaakov, ``{Nonperturbative Tests of
  Three-Dimensional Dualities},''
  \href{http://dx.doi.org/10.1007/JHEP10(2010)013}{{\em JHEP} {\bf 10} (2010)
  013},
\href{http://arxiv.org/abs/1003.5694}{{\tt arXiv:1003.5694 [hep-th]}}.
%%CITATION = ARXIV:1003.5694;%%.

\bibitem{Amariti:2015vwa}
A.~Amariti, ``{Integral identities for 3d dualities with SP(2N) gauge
  groups},''
\href{http://arxiv.org/abs/1509.02199}{{\tt arXiv:1509.02199 [hep-th]}}.
%%CITATION = ARXIV:1509.02199;%%.

\bibitem{Kapustin:2011jm}
A.~Kapustin and B.~Willett, ``{Generalized Superconformal Index for Three
  Dimensional Field Theories},''
\href{http://arxiv.org/abs/1106.2484}{{\tt arXiv:1106.2484 [hep-th]}}.
%%CITATION = ARXIV:1106.2484;%%.

\bibitem{Gahramanov:2013rda}
I.~Gahramanov and H.~Rosengren, ``{A new pentagon identity for the tetrahedron
  index},'' \href{http://dx.doi.org/10.1007/JHEP11(2013)128}{{\em JHEP} {\bf
  11} (2013)  128},
\href{http://arxiv.org/abs/1309.2195}{{\tt arXiv:1309.2195 [hep-th]}}.
%%CITATION = ARXIV:1309.2195;%%.

\bibitem{Gahramanov:2016wxi}
I.~Gahramanov and H.~Rosengren, ``{Basic hypergeometry of supersymmetric
  dualities},'' \href{http://dx.doi.org/10.1016/j.nuclphysb.2016.10.004}{{\em
  Nucl. Phys.} {\bf B913} (2016)  747--768},
\href{http://arxiv.org/abs/1606.08185}{{\tt arXiv:1606.08185 [hep-th]}}.
%%CITATION = ARXIV:1606.08185;%%.

\bibitem{Benini:2011nc}
F.~Benini, T.~Nishioka, and M.~Yamazaki, ``{4d Index to 3d Index and 2d
  TQFT},'' \href{http://dx.doi.org/10.1103/PhysRevD.86.065015}{{\em Phys. Rev.}
  {\bf D86} (2012)  065015},
\href{http://arxiv.org/abs/1109.0283}{{\tt arXiv:1109.0283 [hep-th]}}.
%%CITATION = ARXIV:1109.0283;%%.

\bibitem{Imamura:2012rq}
Y.~Imamura and D.~Yokoyama, ``{$S^3/Z_n$ partition function and dualities},''
  \href{http://dx.doi.org/10.1007/JHEP11(2012)122}{{\em JHEP} {\bf 11} (2012)
  122},
\href{http://arxiv.org/abs/1208.1404}{{\tt arXiv:1208.1404 [hep-th]}}.
%%CITATION = ARXIV:1208.1404;%%.

\bibitem{Imamura:2013qxa}
Y.~Imamura, H.~Matsuno, and D.~Yokoyama, ``{Factorization of the
  $S^3/\mathbb{Z}_n$ partition function},''
  \href{http://dx.doi.org/10.1103/PhysRevD.89.085003}{{\em Phys. Rev.} {\bf
  D89} (2014) no.~8, 085003},
\href{http://arxiv.org/abs/1311.2371}{{\tt arXiv:1311.2371 [hep-th]}}.
%%CITATION = ARXIV:1311.2371;%%.

\bibitem{Aharony:2013dha}
O.~Aharony, S.~S. Razamat, N.~Seiberg, and B.~Willett, ``{3d dualities from 4d
  dualities},'' \href{http://dx.doi.org/10.1007/JHEP07(2013)149}{{\em JHEP}
  {\bf 07} (2013)  149},
\href{http://arxiv.org/abs/1305.3924}{{\tt arXiv:1305.3924 [hep-th]}}.
%%CITATION = ARXIV:1305.3924;%%.

\bibitem{Bozkurt:2018xno}
D.~N. Bozkurt and I.~Gahramanov, ``{Pentagon identities arising in
  supersymmetric gauge theory computations},''
  \href{http://dx.doi.org/10.1134/S0040577919020028}{{\em Teor. Mat. Fiz.} {\bf
  198} (2019) no.~2, 215--224}, \href{http://arxiv.org/abs/1803.00855}{{\tt
  arXiv:1803.00855 [math-ph]}}.
[Theor. Math. Phys.198,no.2,189(2019)].
%%CITATION = ARXIV:1803.00855;%%.

\bibitem{Sarkissian:2018ppc}
G.~Sarkissian and V.~P. Spiridonov, ``{From rarefied elliptic beta integral to
  parafermionic star-triangle relation},''
  \href{http://dx.doi.org/10.1007/JHEP10(2018)097}{{\em JHEP} {\bf 10} (2018)
  097},
\href{http://arxiv.org/abs/1809.00493}{{\tt arXiv:1809.00493 [hep-th]}}.
%%CITATION = ARXIV:1809.00493;%%.

\bibitem{Teschner:2012em}
J.~Teschner and G.~Vartanov, ``{6j symbols for the modular double, quantum
  hyperbolic geometry, and supersymmetric gauge theories},''
  \href{http://dx.doi.org/10.1007/s11005-014-0684-3}{{\em Lett. Math. Phys.}
  {\bf 104} (2014)  527--551},
\href{http://arxiv.org/abs/1202.4698}{{\tt arXiv:1202.4698 [hep-th]}}.
%%CITATION = ARXIV:1202.4698;%%.

\bibitem{Kashaev:2012cz}
R.~Kashaev, F.~Luo, and G.~Vartanov, ``{A TQFT of Turaev\textendash{}Viro Type
  on Shaped Triangulations},''
  \href{http://dx.doi.org/10.1007/s00023-015-0427-8}{{\em Annales Henri
  Poincare} {\bf 17} (2016) no.~5, 1109--1143},
  \href{http://arxiv.org/abs/1210.8393}{{\tt arXiv:1210.8393 [math.QA]}}.

\bibitem{Gahramanov:2014ona}
I.~Gahramanov and H.~Rosengren, ``{Integral pentagon relations for 3d
  superconformal indices},'' \href{http://arxiv.org/abs/1412.2926}{{\tt
  arXiv:1412.2926 [hep-th]}}.
[Proc. Symp. Pure Math.93,165(2016)].
%%CITATION = ARXIV:1412.2926;%%.

\bibitem{pachner1991pl}
U.~Pachner, ``P.l. homeomorphic manifolds are equivalent by elementary
  shellings,'' \href{http://dx.doi.org/10.1016/S0195-6698(13)80080-7}{{\em
  European journal of Combinatorics} {\bf 12} (1991) no.~2, 129--145}.

\bibitem{Dimofte:2011ju}
T.~Dimofte, D.~Gaiotto, and S.~Gukov, ``{Gauge Theories Labelled by
  Three-Manifolds},'' \href{http://dx.doi.org/10.1007/s00220-013-1863-2}{{\em
  Commun. Math. Phys.} {\bf 325} (2014)  367--419},
  \href{http://arxiv.org/abs/1108.4389}{{\tt arXiv:1108.4389 [hep-th]}}.

\bibitem{Benvenuti:2016wet}
S.~Benvenuti and S.~Pasquetti, ``{3d $ \mathcal{N} $ = 2 mirror symmetry,
  pq-webs and monopole superpotentials},''
  \href{http://dx.doi.org/10.1007/JHEP08(2016)136}{{\em JHEP} {\bf 08} (2016)
  136}, \href{http://arxiv.org/abs/1605.02675}{{\tt arXiv:1605.02675
  [hep-th]}}.

\bibitem{Jafarzade:2018yei}
S.~Jafarzade, ``{New Pentagon Identities Revisited},''
  \href{http://dx.doi.org/10.1088/1742-6596/1194/1/012054}{{\em J. Phys. Conf.
  Ser.} {\bf 1194} (2019) no.~1, 012054},
\href{http://arxiv.org/abs/1812.01325}{{\tt arXiv:1812.01325 [math-ph]}}.
%%CITATION = ARXIV:1812.01325;%%.

\bibitem{Baxter:1987eq}
R.~J. Baxter, J.~H.~H. Perk, and H.~Au-Yang, ``{New solutions of the star
  triangle relations for the chiral Potts model},''
\href{http://dx.doi.org/10.1016/0375-9601(88)90896-1}{{\em Phys. Lett.} {\bf
  A128} (1988)  138--142}.
%%CITATION = PHLTA,A128,138;%%.

\bibitem{perk1986nonintersecting}
J.~Perk and F.~Wu, ``Nonintersecting string model and graphical approach:
  Equivalence with a potts model,'' {\em Journal of statistical physics} {\bf
  42} (1986) no.~5-6, 727--742.

\bibitem{Yamazaki:2012cp}
M.~Yamazaki, ``{Quivers, YBE and 3-manifolds},''
  \href{http://dx.doi.org/10.1007/JHEP05(2012)147}{{\em JHEP} {\bf 05} (2012)
  147},
\href{http://arxiv.org/abs/1203.5784}{{\tt arXiv:1203.5784 [hep-th]}}.
%%CITATION = ARXIV:1203.5784;%%.

\bibitem{Gahramanov:2017idz}
I.~Gahramanov and S.~Jafarzade, ``{Comments on the multi-spin solution to the
  Yang-Baxter equation and basic hypergeometric sum/integral identity},''
  \href{http://dx.doi.org/10.1142/S0217732319501402}{{\em {Mod. Phys. Lett.}}
  {\bf A34,no.18,1950140} (2019)  },
\href{http://arxiv.org/abs/1710.09106}{{\tt arXiv:1710.09106 [math-ph]}}.
%%CITATION = ARXIV:1710.09106;%%.

\bibitem{Yamazaki:2015voa}
M.~Yamazaki and W.~Yan, ``{Integrability from 2d ${\mathcal{N}}=(2,2)$
  dualities},'' \href{http://dx.doi.org/10.1088/1751-8113/48/39/394001}{{\em J.
  Phys.} {\bf A48} (2015)  394001},
\href{http://arxiv.org/abs/1504.05540}{{\tt arXiv:1504.05540 [hep-th]}}.
%%CITATION = ARXIV:1504.05540;%%.

\bibitem{He:2004rn}
Y.-H. He, ``{Lectures on D-branes, gauge theories and Calabi-Yau
  singularities},'' in {\em {1st Hangzhou-Beijing International Summer School
  Beijing, China, August 6, 2004}}.
\newblock 2004.
\newblock
\href{http://arxiv.org/abs/hep-th/0408142}{{\tt arXiv:hep-th/0408142
  [hep-th]}}.
\newblock
%%CITATION = HEP-TH/0408142;%%.

\bibitem{Yamazaki:2008bt}
M.~Yamazaki, ``{Brane Tilings and Their Applications},''
  \href{http://dx.doi.org/10.1002/prop.200810536}{{\em Fortsch. Phys.} {\bf 56}
  (2008)  555--686},
\href{http://arxiv.org/abs/0803.4474}{{\tt arXiv:0803.4474 [hep-th]}}.
%%CITATION = ARXIV:0803.4474;%%.

\bibitem{Spiridonov:2019uuw}
V.~P. Spiridonov, ``{The rarefied elliptic Bailey lemma and the
  Yang\textendash{}Baxter equation},''
  \href{http://dx.doi.org/10.1088/1751-8121/ab3358}{{\em J. Phys. A} {\bf 52}
  (2019) no.~35, 355201}, \href{http://arxiv.org/abs/1904.12046}{{\tt
  arXiv:1904.12046 [math-ph]}}.

\bibitem{Bult2007}
F.~J. van~de Bult, E.~M. Rains, and J.~V. Stokman, ``Properties of generalized
  univariate hypergeometric functions,'' {\em Communications in Mathematical
  Physics} {\bf 275} (2007) no.~1, 37--95.

\bibitem{Dimofte:2012pd}
T.~Dimofte and D.~Gaiotto, ``{An E7 Surprise},''
  \href{http://dx.doi.org/10.1007/JHEP10(2012)129}{{\em JHEP} {\bf 10} (2012)
  129}, \href{http://arxiv.org/abs/1209.1404}{{\tt arXiv:1209.1404 [hep-th]}}.

\bibitem{Eren:2019ibl}
E.~Eren, I.~Gahramanov, S.~Jafarzade, and G.~Mogol, ``{Gamma function solutions
  to the star-triangle equation},''
  \href{http://dx.doi.org/10.1016/j.nuclphysb.2020.115283}{{\em Nucl. Phys. B}
  {\bf 963} (2021)  115283}, \href{http://arxiv.org/abs/1912.12271}{{\tt
  arXiv:1912.12271 [math-ph]}}.

\bibitem{Derkachov:2019ynh}
S.~Derkachov and A.~N. Manashov, ``{On complex Gamma function integrals},''
\href{http://arxiv.org/abs/1908.01530}{{\tt arXiv:1908.01530 [math-ph]}}.
%%CITATION = ARXIV:1908.01530;%%.

\bibitem{Ruijsenaars2003}
S.~Ruijsenaars, ``A generalized hypergeometric function iii. associated hilbert
  space transform,''
  \href{http://dx.doi.org/https://doi.org/10.1007/s00220-003-0970-x}{{\em
  Communications in Mathematical Physics} {\bf 243} (2003) no.~3, 413--448}.

\bibitem{STOKMAN2005119}
J.~V. Stokman, ``Hyperbolic beta integrals,''
  \href{http://dx.doi.org/http://dx.doi.org/10.1016/j.aim.2003.12.003}{{\em
  Advances in Mathematics} {\bf 190} (2005) no.~1, 119 -- 160}.

\bibitem{Kels:2018xge}
A.~P. Kels, ``{Integrable quad equations derived from the quantum Yang-Baxter
  equation},''
\href{http://arxiv.org/abs/1803.03219}{{\tt arXiv:1803.03219 [math-ph]}}.
%%CITATION = ARXIV:1803.03219;%%.

\bibitem{Gahramanov:2021pgu}
I.~Gahramanov and O.~E. Kaluc, ``{Bailey pairs for the q-hypergeometric
  integral pentagon identity},'' \href{http://arxiv.org/abs/2111.14793}{{\tt
  arXiv:2111.14793 [math-ph]}}.

\end{thebibliography}\endgroup

%\providecommand{\href}[2]{#2}\begingroup\raggedright\begin{thebibliography}{100}

%\end{thebibliography}\endgroup

\end{document}